\documentclass{jfm}
\usepackage{amssymb}
\usepackage{natbib}
\usepackage{lineno}
\usepackage{amsfonts,amsbsy}
\usepackage{graphicx}
\usepackage{color}

\linenumbers

\def\e{\mathrm{e}}
\def\i{\mathrm{i}}
\def\d{\mathrm{d}}

\def\beq{\begin{equation}}
\def\eeq{\end{equation}}

\def\Ai{\mathrm{Ai}}

\def\bx{\boldsymbol{x}}
\def\bk{\boldsymbol{k}}
\def\bu{\boldsymbol{u}}

\def\bxi{\boldsymbol{\xi}}

\def\bW{\boldsymbol{W}}

\def\bchi{\boldsymbol{\chi}}

\def\bq{\boldsymbol{q}}
\def\bX{\boldsymbol{X}}
\def\Pe{\mathrm{Pe}}
\def\E{\mathbb{E}\,}

\newcommand{\av}[1]{\langle #1 \rangle}

\newcommand{\dt}[2]{\frac{\mathrm{d} #1}{\mathrm{d} #2}}

\newcommand{\eqn}[1]{(\ref{eqn:#1})}
\newcommand{\lab}[1]{\label{eqn:#1}}
\newcommand{\inter}[1]{\quad \textrm{#1} \quad}

\def\keff{\mathsf{k}}

\def\XXint#1#2#3{{\setbox0=\hbox{$#1{#2#3}{\int}$}
\vcenter{\hbox{$#2#3$}}\kern-.5\wd0}}

\title[Dispersion in the large-deviation regime I]{Dispersion in the large-deviation regime. Part I: shear flows and periodic flows}
\author[P. H. Haynes and J. Vanneste]{P. H. Haynes$^1$ and J. Vanneste$^2$\thanks{Email address for correspondence: J.Vanneste@ed.ac.uk}}
\affiliation{$^1$Department of Applied Mathematics and Theoretical Physics, University of Cambridge, Wilberforce Road, Cambridge CB3 0WA, UK \\[\affilskip] 
$^2$School of Mathematics and Maxwell Institute for Mathematical Sciences, University of Edinburgh, King's Buildings, Edinburgh EH9 3JZ, UK}

\begin{document}

\maketitle

\begin{abstract}
The dispersion of a passive scalar in a fluid through the combined action of advection and molecular diffusion is often described as a diffusive process, with an effective diffusivity that is enhanced  compared to the molecular value. However, this description fails to capture the tails of the scalar concentration distribution in initial-value problems. To remedy this, we develop a large-deviation theory of scalar dispersion that provides an approximation to the scalar concentration 
valid at much larger distances away from the centre of mass, specifically distances that are $O(t)$ rather than $O(t^{1/2})$, where $t \gg 1$ is the time from the scalar release. 

The theory centres on the calculation of a rate function characterising the large-time form of the scalar concentration. This function is deduced from the solution of a one-parameter family of eigenvalue problems which we derive using two alternative approaches, one asymptotic, the other probabilistic. We emphasise the connection between the large-deviation theory and the homogenisation theory that is often used to compute effective diffusivities:  a perturbative solution of the eigenvalue problems in the appropriate limit reduces at leading order to the cell problem of homogenisation theory.

We consider two classes of flows in some detail: shear flows and periodic flows with closed streamlines (cellular flows). In both cases, large deviation generalises classical results on effective diffusivity and captures new phenomena relevant to the tails of the scalar distribution. These include approximately finite dispersion speeds arising at large P\'eclet number $\Pe$ (corresponding to small molecular diffusivity) and,  for  two-dimensional cellular flows, anisotropic dispersion. Explicit asymptotic results are obtained for shear flows in the limit of large  $\Pe$. (A companion paper, Part II, is devoted to the large-$\Pe$ asymptotic treatment of cellular flows.) The predictions of large-deviation theory are compared with Monte Carlo simulations that estimate the tails of concentration accurately using importance sampling.

\end{abstract}

\section{Introduction}

\citet{tayl53} identified the phenomenon of shear dispersion in which a
passive scalar, e.g.\ a chemical pollutant, released in a pipe Poiseuille flow spreads along the pipe according to a diffusion law. The corresponding diffusivity, often termed effective diffusivity to distinguish it from molecular diffusivity, is inversely proportional to  molecular diffusivity when the latter is small \citep[see also][]{aris56,youn-jone}. This effective diffusivity is associated with a random walk along the pipe that results from the random sampling of the Poiseuille flow by  molecular Brownian motion across the pipe. The diffusive description of this random walk, and the corresponding Gaussian profile of the scalar concentration, of course only apply on time scales that are much longer than the Lagrangian correlation time scale.

Shear dispersion is a striking example of a broad class of phenomena in which the interaction between fluid motion and Brownian motion leads to a strong enhancement of dispersion and to effective diffusivities that are orders of magnitude larger than molecular diffusivity. The importance of these phenomena in applications, in particular industrial, biological and environmental applications, is obvious. 
This has motivated studies of  effective diffusivity in many different flows \citep[see][for a review]{majd-kram}. These include spatially periodic flows  which can be analysed using the method of homogenisation. This method, which exploits the separation between the (small) scale of the flow  and the (large) scale of the scalar field that emerges in the long-time limit, has proved highly valuable: it applies to  more complicated flows, including time-dependent and random flows, and provides a unifying framework for methods used earlier. Shear dispersion, in particular, can be regarded as a special case of homogenisation applied to periodic flows, where cells repeat in the along pipe direction and the flow in each cell is simple
Poiseuille flow. 

In the large literature on shear dispersion, efforts have been made to overcome the restriction to large times that underlies the diffusive approximation, and improved asymptotic estimates that capture some of the early-time behaviour have been obtained (see \citealt{youn-jone} for a review and \citealt{cama-et-al} for more recent results). For periodic flows, because the effective diffusivity  is more difficult to compute,  the focus has  mainly  remained on the derivation of asymptotic estimates and bounds, in particular in the limit of small molecular diffusivity \citep[e.g.][]{majd-kram,novi-et-al}.

Here we consider a different aspect. The characterisation of dispersion in the long-time limit $t \gg 1$ by an effective diffusivity and hence by a Gaussian scalar distribution holds only close to the centre of mass of the distribution: the results of homogenisation are in essence a manifestation of the central-limit theorem and apply only to particles displaced from the mean by $O(t^{1/2})$ distances. 
Our aim is to go beyond this and describe the concentration far from the mean. To achieve this, 
we derive large-deviation estimates for the concentration, that is, we derive the  rate function $g$ in an approximation of the form $\exp(-tg(\bx/t))$ for the scalar concentration at position $\bx$ and time $t$.

\begin{figure}
\begin{center}
\includegraphics[height=6cm]{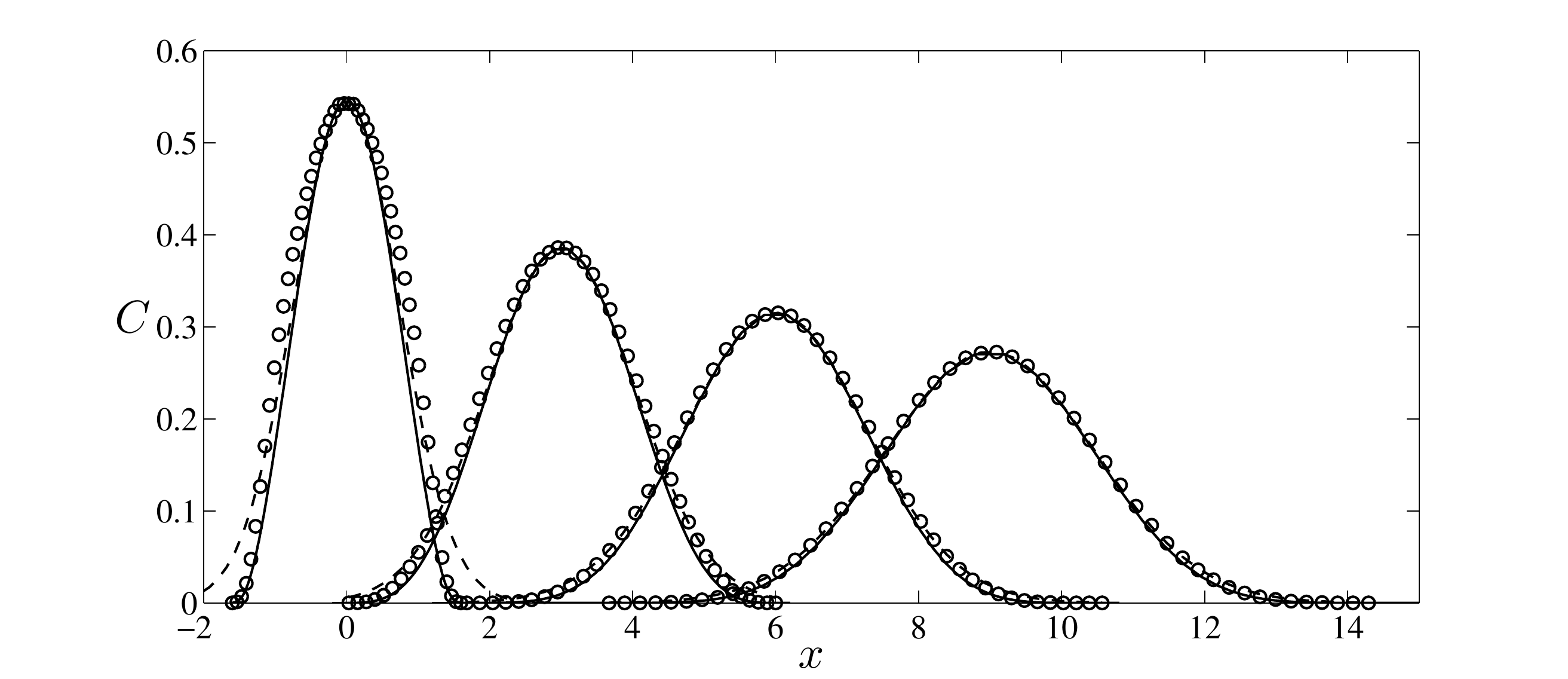} \\
\includegraphics[height=6cm]{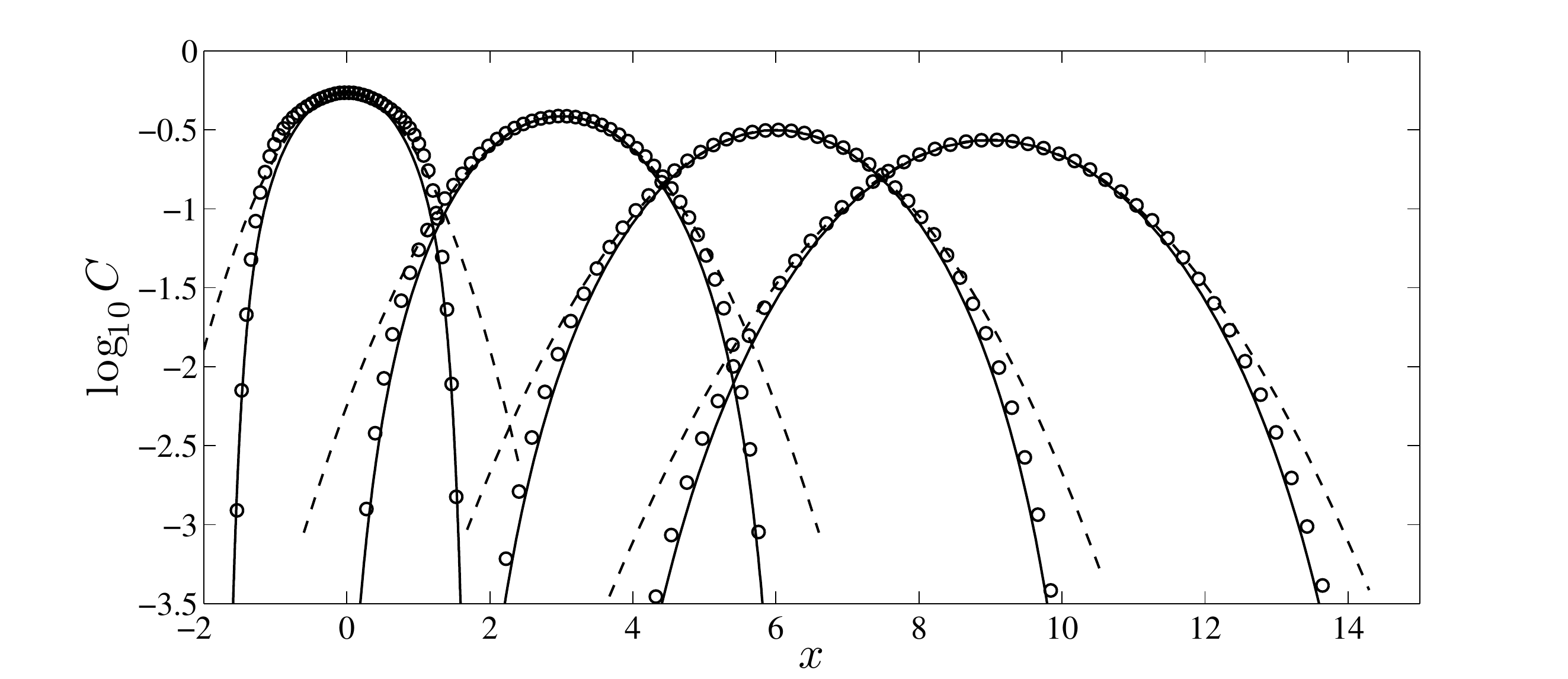}
\caption{Cross-section averaged concentration $C(x,t)$ (top panel) and its logarithm $\log_{10} C(x,t)$ (bottom panel) in a Couette flow as a function of $x$ for $t=2,\, 4,\, 6$ and $8$ (from left to right, curves have been offset for clarity). Monte Carlo results (symbols) are compared with the large-deviation and diffusive predictions (solid and dashed lines).} \label{fig:pdfcouette}
\end{center}
\end{figure}

Large-deviation theory extends the central-limit theorem and applies to numerous probabilistic problems \citep[e.g.][]{demb-zeit,denh00}. When applied to the stochastic differential equations governing the motion of fluid particles advected and diffused in a fluid flow, it naturally yields an improved approximation to the scalar concentration (interpreted as a particle-position probability function, cf.\ \citealt{jans-roge}). This approximation is valid for distances from the mean that are $O(t)$ rather than $O(t^{1/2})$ and therefore captures the tails of the distribution. These are typically non-Gaussian and not adequately represented by the diffusive approximation. This  is illustrated in Figure \ref{fig:pdfcouette} by the example of dispersion in a plane Couette flow, one of the shear flows considered in detail in this paper. The top panel shows the profile along the flow of the cross-stream averaged concentration $C(x,t)$ at four successive times in the case of small molecular diffusivity. The figure compares the averaged concentration obtained numerically using a Monte Carlo simulation (symbols) with the Gaussian, diffusive approximation (dashed lines) and the large-deviation approximation derived in \S\S2--3 (solid lines). The units of $x$ and $t$ have been chosen so that the maximum flow velocity and (Taylor) effective diffusivity are both $1$. The inadequacy of the diffusive approximation in describing the tails of the concentration and the superiority of the large-deviation approximation are apparent in the top panel for the earliest profile $C(x,t=2)$. They are obvious for all the profiles in the bottom panel which displays the results using logarithmic scale for $C(x,t)$. This emphasises the tails of $C(x,t)$ to reveal how the diffusive prediction overestimates dispersion and to demonstrate the effectiveness of the large-deviation approximation. We note that while large deviation formally applies for $t \gg 1$, it appears here remarkably accurate for moderate $t$. (The discrepancies  between large-deviation and Monte Carlo results for $t \ge 4$ are mainly attributable to the limitations of the straightforward Monte Carlo method used here and are much reduced with the more sophisticated methods  discussed in \S3.)

As the Couette-flow example illustrates, large-deviation theory provides estimates of the low scalar concentrations in the tails, where the diffusive approximation fails. This makes it relevant to a range of applications in which low concentrations matter. Examples include the prediction of the first time at which the concentration of a pollutant released in the environment exceeds a low safety threshold, and the quantification of the impact of stirring on chemical reactions in a fluid. In such examples, there is a strong sensitivity of the response (physiological or chemical) to low scalar concentrations that makes the logarithm of the concentration, and hence the rate function $g$, highly relevant quantities. This broad observation can be made  precise for the certain classes of chemical reactions. For F-KPP reactions \citep[e.g.][]{xin09}, the combination of  diffusion and reaction leads to the formation of concentration fronts that propagate at a speed that turns out to be controlled by the large-deviation statistics of the dispersion and given explicitly in terms of the rate function $g$ (\citealt{gart-frei}; see also \citealt{frei85}, Ch.\ 7, \citealt{xin09}, Ch.\ 2, and \citealt{tzel-v14b}). 

The present paper starts in \S\ref{sec:formulation} with a relatively general treatment of the large-deviation theory of dispersion  which applies to time-independent periodic flows and to shear flows. The key result is a family of eigenvalue problems parameterised by a variable $\bq$.
The principal eigenvalue, $f(\bq)$, is the Legendre transform of the rate function $g$. These eigenvalue problems can be thought of as generalised cell problems in that they resemble and extend the cell problem that appears when homegenization is used to compute effective diffusivities. In \S\S\ref{sec:ldev}--\ref{sec:prob} we present two alternative derivations of the the eigenvalue problems: the first is a direct asymptotic method that treats the  large-deviation form of the concentration as an ansatz 
\citep[see][]{kusk-kell}; the second follows the standard probabilistic approach based on the Ellis--G\"artner theorem and considers the cumulant generating function of the particle position \citep[e.g.][]{elli95,demb-zeit,denh00,touc09}. We then discuss the relation between large deviation and homogenisation (\S\ref{sec:hom}).  Homogenisation, and the corresponding diffusive approximation, are shown to be recovered when the eigenvalue problems yielding $f(\bq)$ are solved perturbatively for small $|\bq|$ up to $O(|\bq|^3)$ errors. Carrying out the perturbation expansion to higher orders provides a systematic way of improving on the diffusive approximation; in the case of shear dispersion, this recovers earlier results \citep{merc-robe, youn-jone}. 

The rest of the paper is devoted to dispersion in specific shear and periodic flows. We compute the functions $f$ and $g$ for the classical Couette and Poiseuille flows in \S\ref{sec:shear} by solving the relevant one-dimensional eigenvalue problem numerically. We also obtain asymptotic results for the concentration at small and large distances from the centre of mass. While the first limit recovers the well-known expression for the effective diffusivity of shear flows, the second captures the finite propagation speed that exists when diffusion along the pipe is neglected. This provides a transparent example of the limitations of the diffusive approximation. Section \ref{sec:cell} is devoted to a standard example of periodic flow, the two-dimensional cellular flow with streamfunction $\psi = - \sin x \sin y$. The numerical solution of the corresponding eigenvalue problems for specific values of the P\'eclet number $\Pe$ (measuring the relative strength of advection and diffusion) reveals interesting features of the dispersion, such as  anisotropy, that are not captured in the diffusive approximation. Using a regular perturbation expansion, we derive explicit results in the limit of small $\Pe$. We examine the opposite, large-P\'eclet-number limit in a companion paper \citep[][hereafter Part II]{hayn-v14b}. We conclude the paper with a Discussion in \S\ref{sec:disc}.

Throughout the present paper and Part II, we verify the predictions of large-deviation theory against direct Monte Carlo simulations of particle dispersion. This is not without challenges since this requires estimating the tails of distributions which are associated with rare events and are, by definition, difficult to sample. We have therefore used importance sampling and implemented two methods that are applicable broadly. These are described in Appendix \ref{sec:motecarlo}. Two other Appendices are devoted to technical details of certain asymptotic limits.

\section{Formulation} \label{sec:formulation}

We start with the advection--diffusion equation for the concentration $C(\bx,t)$ of a passive scalar. Using a characteristic spatial scale $a$ as reference length and the corresponding diffusive time scale $a^2/\kappa$, where $\kappa$ is the molecular diffusivity, as a reference time, this equation can be written in the non-dimensional form
\beq \lab{ad-dif}
\partial_t C + \Pe \, \bu \cdot \nabla C = \nabla^2 C,
\eeq
where $\Pe = U a/\kappa$ is the P\'eclet number. Here $U$ is the typical magnitude of the velocity field, which is assumed to be time independent, $\bu=\bu(\bx)$, and divergence free, $\nabla \cdot \bu =0$. 

Equation \eqn{ad-dif} can be considered as the Fokker--Planck equation associated with the stochastic differential equation (SDE) which governs the position of fluid particles, 
\beq \lab{sde}
\d \bX = \Pe \, \bu(\bX) \d t + \sqrt{2} \, \d \bW,
\eeq
where $\bW$ denotes a Brownian motion. In this interpretation and with $\bX(0)=\bx_0$, the initial condition for the concentration is $C(\bx,0)=\delta(\bx-\bx_0)$ and the concentration at later times can then be thought of as the transition probability for a particle to move from $\bx_0$ at $t=0$ to $\bx$ at $t$. We focus on this initial condition and use the  notation $C(\bx,t|\bx_0)$ when the dependence on $\bx_0$ needs to be made explicit.

In this paper we consider two somewhat different flow configurations. The first, relevant to Taylor dispersion, corresponds to parallel shear flows, with $\bu(\bx)$ unidirectional and varying in the cross-flow direction only, and a domain that is bounded in this direction. The concentration $C(\bx,t|\bx_0)$ then satisfies a no-flux condition at the boundary. The second configuration corresponds to a periodic $\bu(\bx)$ in an unbounded domain. In both cases, our interest is in the dispersion in the unbounded directions of the domain. The shear-flow configuration can essentially be regarded as a particular case of the more general periodic-flow configuration, with the domain extending over only one period in the streamwise direction and no-flux boundary conditions replacing periodicity conditions. Because of this, we consider the two configurations together when developing the general large-deviation approach in the rest of this section. Any ambiguity that may arise as a result will be clarified in \S\ref{sec:shear} and \S\ref{sec:cell} when we apply the approach separately to  shear flows and to two-dimensional periodic flows and obtain explicit results. Mixed configurations, in which the flow is periodic in certain directions and bounded in others, could also be treated with no essential changes. 

\subsection{Large-deviation approximation}
\label{sec:ldev}

We are interested in the form of $C(\bx,t|\bx_0)$ for $t \gg 1$. Under the assumption that $|\bx-\bx_0|/t = O(1)$, the solution to \eqn{ad-dif} can be sought as the expansion
\beq \lab{expC}
C(\bx,t|\bx_0) = t^{-d/2} \e^{-t g(\bxi)} \left( \phi_0(\bx,\bxi) + t^{-1} \phi_1(\bx,\bxi) + \cdots \right), \quad \textrm{where} \ \ \bxi = (\bx-\bx_0)/t,
\eeq
where $d$ is the number of spatial dimensions. This can be considered to be a WKB expansion with $t$ as large parameter. The leading-order approximation
\beq \lab{largedevi}
C(\bx,t|\bx_0) \sim t^{-d/2} \phi(\bx,\bxi) \e^{-t g(\bxi)}, 
\eeq
has  the characteristic large-deviation form in which  $g(\bxi)$ is the Cram\'er or rate function \citep[e.g.][and references therein]{demb-zeit,touc09}. The conservation of total mass -- the spatial integral of $C(\bx,t|\bx_0)$ -- imposes that
\beq \lab{g0}
\min_{\bxi} g(\bxi) = 0
\eeq
and explains the presence of the prefactor $t^{-d/2}$ in \eqn{largedevi}, as an application of Laplace's method shows. Note that we concentrate on this leading-order approximation throughout and hence omit the subscript $0$ from $\phi$.

 Introducing the expansion \eqn{expC} into \eqn{ad-dif} and retaining only the leading order terms  gives
\beq
(\bxi \cdot \nabla_{\bxi} g - g) \phi = \nabla^2 \phi - \left(\Pe \, \bu + 2 \nabla_{\bxi} g\right) \cdot \nabla \phi + \left(\Pe \, \bu \cdot \nabla_{\bxi} g + | \nabla_{\bxi} g |^2\right) \phi.
\eeq
Letting
\beq \lab{legendre}
\bq = \nabla_{\bxi} g \inter{and} f(\bq) = \bq \cdot {\bxi}  - g,
\eeq
this equation reduces to
\beq \lab{eig1}
 \nabla^2 \phi - \left(\Pe \, \bu + 2 \bq \right) \cdot \nabla \phi + \left(\Pe \,\bu \cdot \bq + |\bq|^2\right) \phi = f(\bq) \phi,
\eeq
where $\bq$ can be regarded as a parameter. This can be rewritten compactly as 
\beq \lab{expq}
\e^{\bq \cdot \bx} \left(\nabla^2 - \Pe \, \bu \cdot \nabla \right) \left(\e^{-\bq \cdot \bx} \phi \right) = f(\bq) \phi,
\eeq
in which the form of the operator on the left-hand side 
makes transparent the connection to the advection--diffusion operator $\nabla^2 - \Pe \, \bu \cdot \nabla$. 
The function $\phi$ satisfies no-flux boundary conditions when impermeable boundaries are present or periodic boundary conditions in the case of unbounded domains with periodic $\bu(\bx)$.

Equation \eqn{eig1} is central to this paper. Together with its associated boundary conditions, it gives a family of eigenvalue problems for $\phi$ parameterised by $\bq$, with $f(\bq)$ as the eigenvalue. Solving these eigenvalue problems (numerically in general) provides $f(\bq)$ as the principal eigenvalue, that is, the eigenvalue with largest real part. The rate function $g(\bxi)$ is then recovered by noting from \eqn{legendre} that $g(\bxi)$ and $f(\bq)$ are related by a Legendre transform
\beq \lab{legendre1}
f(\bq) = \sup_{\bxi} \left(\bq \cdot \bxi - g(\bxi)\right) \inter{and}
g(\bxi)= \sup_{\bq} \left(\bxi \cdot \bq - f(\bq)\right).
\eeq
The fact that the critical points of $f$ are suprema and the convexity of $f$ can be deduced from the probabilistic interpretation of $f(\bq)$ discussed  below.\footnote{Note that the second equality assumes that $f$ is differentiable \citep[e.g.][]{touc09}.} It follows that
\beq \lab{qxi}
\bxi = \nabla_{\bq} f,
\eeq
which gives a one-to-one map between the parameter $\bq$ and the physical variable $\bxi=\bx/t$. The eigenfunction $\phi$ of \eqn{eig1} associated with $f(\bq)$ can therefore be equivalently thought of as a function of $\bxi$, as in \eqn{largedevi}, or of $\bq$, as in \eqn{eig1}. Note that the maximum principle can be used to show that $f(\bq)$ is real and that $\phi$ is sign definite  \citep[e.g.][]{bere-et-al}. This is consistent with the asymptotics \eqn{largedevi} and the observation that the concentration $C(\bx,t|\bx_0)$ is positive for all time if it is initially positive.

To summarise,  solving the eigenvalue problem \eqn{eig1} for arbitrary $\bq$ and performing a Legendre transform of the principal eigenvalue yields the large-$t$ approximation \eqn{largedevi} of the concentration. This approximation is valid for $|\bx|=O(t)$ and thus, as discussed below, extends the standard diffusive approximation which requires $|\bx|=O(t^{1/2})$. The eigenvalue problem \eqn{eig1} can be thought of as a generalised cell problem since, as discussed in \S\,\ref{sec:hom}, it generalises the cell problem of homogenisation theory. \citet[][\S4.3.1]{bens-et-al} derive this eigenvalue problem as part of a Floquet--Bloch theory for linear equations with periodic coefficients and term it `shifted cell problem' (see also \citealt[][\S3.6]{papa95}, and \S4 below).


\subsection{Probabilistic derivation} \label{sec:prob}

An alternative view of the problem considers the moment generating function
\beq \lab{genfun}
w(\bq,\bx,t)= \E \e^{\bq \cdot \bX}, \quad \textrm{with} \ \ \bX(0)=\bx
\eeq
for the position of the fluid particles satisfying \eqn{sde}. Here $\E$ denotes the expectation over the Brownian process in \eqn{sde}. The generating function obeys the backward Kolmogorov equation
\beq \lab{backkol}
\partial_t w = \Pe \, \bu \cdot \nabla w + \nabla^2 w, \quad \textrm{with} \ \ w(\bq,\bx,0)=\e^{\bq \cdot \bx}
\eeq
\citep[e.g.][]{okse98,gard04}.
A solution can be sought in the form
\beq \lab{w}
w(\bq,\bx,t)=\e^{\bq \cdot \bx + f(\bq) t} \phi^\dagger(\bq,\bx),
\eeq
where the function $f(\bq)$ remains to be determined but will shortly be identified with that in \eqn{legendre}. 

Introducing \eqn{w} into \eqn{backkol} leads to
\beq \lab{eig2}
\nabla^2 \phi^\dagger + \left(\Pe \, \bu + 2 \bq \right) \cdot \nabla \phi^\dagger + \left( \Pe \, \bu \cdot \bq + | \bq |^2 \right) \phi^\dagger = f(\bq) \phi^\dagger,
\eeq
with no-flux or periodic boundary conditions.
This corresponds to a family of eigenvalue problems, again parameterised by $\bq$, which are the adjoints of those in \eqn{eig1}, and hence have the same eigenvalues and in particular the same principal eigenvalue $f(\bq)$, justifying the notation in \eqn{w}.  This eigenvalue controls $w(\bx,t)$ for $t \gg 1$. As a result, it can alternatively  be defined by
\beq \lab{f}
f(\bq) = \lim_{t \to \infty} \frac{1}{t} \log \E \e^{\bq \cdot \bX(t)}
\eeq
and interpreted as the limit as $t \to \infty$ of the cumulant generating function scaled by $t^{-1}$. This function is convex by definition.

The relationship between the large-$t$ asymptotics of $C(\bx,t|\bx_0)$ encoded in $g(\bxi)$ and that of $w(\bx,t)$ can be made obvious. Noting from the definition \eqn{genfun} that $w(\bx,t)$ is the Legendre transform with respect to $\bx'$ of $C(\bx',t|\bx)$ with $-\bq$ the variable dual to $\bx'$, we apply Laplace's method to obtain
\[
w(\bq,\bx,t) = \int \e^{\bq \cdot \bx'} C(\bx',t|\bx) \, \d \bx' \asymp \int \e^{t (\bq \cdot (\bxi+\bx/t) - g(\bxi))} \, \d \bxi \asymp \e^{\bq \cdot \bx + t \sup_{\bxi} (\bq \cdot \bxi - g(\bxi))},
\]
where $\asymp$ denotes the asymptotic equivalence of the logarithms as $t \to \infty$ and we use    \eqn{largedevi} to write $C(\bx',t|\bx) \asymp \exp(-tg((\bx'-\bx)/t))$.

From \eqn{w} we obtain the first part of \eqn{legendre1}. Under the assumption of differentiability of $f(\bq)$, which ensures that $g(\bxi)$ is convex, the second part follows, allowing the computation of the rate function. The argument used in this subsection, which relies on  Laplace's method to establish a connection between rate function $g(\bxi)$ and scaled cumulant generating function $f(\bq)$, is an instance of the G\"artner--Ellis theorem, a fundamental result of large-deviation theory which extends Cram\'er's treatment of the sum of independent random numbers \citep[see, e.g.,][]{elli95,demb-zeit,touc09}. Rigorous results for a problem very similar to that defined above can be found in \citet[][Ch.\ 7]{frei85}. It may be worth contrasting the large-time ($t \gg 1$) large deviations  discussed in this paper, with the small-noise ($\Pe \gg 1$) large deviations developed by Freidlin \& Wentzell \citep[see][]{frei-went12}: while for small noise a single (maximum-likelihood or instanton) trajectory controls the rate function $g$, this is not generally the case for large time. As we discuss in the case of shear flows in \S3, it is only for $\Pe \gg 1$ and $|\bq|$ sufficiently large that $g$ can be expressed in terms of single trajectory and that the two forms of large deviations intersect.

Some properties of $f(\bq)$ and $g(\bxi)$ are useful to infer properties of the dispersion directly from $f(\bq)$ without the need to carry out the Legendre transform explicity. As noted, $f(\bq)$ and $g(\bxi)$ are convex. Therefore, from \eqn{qxi}, increasing $\bq$ correspond to increasing $\bxi$, and $\bq$ can be thought of as a proxy for the more physical variable $\bxi$. 
It is clear from  \eqn{f} that $f(0)=0$; correspondingly,
\beq
\nabla_{\bq} f(0)=\bxi_*,
\eeq 
defines $\bxi_*$ which, by \eqn{legendre1}, minimizes $g$. Eq.\ \eqn{largedevi} then indicates that the maximum of $C(\bx,t)$ and its centre of mass are located at $\bx \sim \bxi_\star t$. Qualitatively the Legendre transform implies that a slow growth of $f(\bq)$ away from its minimum corresponds to a rapid growth of $g(\bxi)$ and vice versa. In particular, linear asymptotes for $f(\bq)$, say $f(q) \sim \lambda q$ as $q \to \infty$ in the one-dimensional case, correspond to vertical asymptotes for $g(\xi)$, $g(\xi) \to \infty$ as $\xi \to \lambda^-$.  This implies that $C(\bx,t)$ vanishes for $x > \lambda_t$, reflecting a finite maximum transport speed for the scalar. 
Exactly linear asymptotes do not arise for $f(\bq)$ because the eigenvalue problem \eqn{eig1} for $|\bq| \gg 1$ has the simple solution $f(\bq)\sim |\bq|^2$ which corresponds to a purely diffusive behaviour. However, for large $\Pe$, there can be a range of values of $\bq$ for which $f(\bq)$ is approximately linear and a finite transport speed controls scalar dispersion.

\subsection{Relation with homogenisation and its extensions} \label{sec:hom}

Much of the literature on scalar dispersion focuses on the computation of an effective diffusivity governing the dispersion for $t \gg 1$ and $|\bx-\bx_0|=O(t^{1/2})$. In this approximation, \eqn{ad-dif} reduces to the diffusion equation
\beq \lab{effdiff}
\partial_t C + \Pe \av{\bu} \cdot \nabla C  = \nabla \cdot \left(\keff \cdot \nabla C\right),
\eeq
where $\av{\bu}$ is the spatial average of $\bu(\bx)$, and $\keff$ is an effective diffusivity tensor. 
Alternatively, $\av{\bu}$ and $\keff$ can be obtained from the particle statistics using
\beq \lab{effdiff2}
\lim_{t \to \infty} \frac{1}{t} \E \bX = \Pe \av{\bu} \inter{and}
\lim_{t \to \infty} \frac{1}{2 t} \E (\bX - \Pe \av{\bu} t) \otimes (\bX- \Pe \av{\bu}t) = \keff.
\eeq
The form of $\keff$ has been derived for a variety of flows using several essentially equivalent methods, starting with Taylor's \citeyearpar{tayl53} work on shear flows. In the last 20 years, homogenisation, as reviewed in \citet{majd-kram} and \citet{pavl-stua}, has become the systematic method of choice. 

The diffusive approximation \eqn{effdiff} can be recovered from the more general large deviation results: since the assumption $|\bx-\bx_0-\Pe \av{\bu}t|=O(t^{1/2})$ implies that $\bxi \ll 1$ and hence that $\bq \ll 1$, we can expand $f(\bq)$ according to
\beq
f(\bq)= \bxi_* \cdot \bq + \frac{1}{2} \bq \cdot \mathsf{H}_f \cdot \bq + O(|\bq|^3),
\eeq
where  $\mathsf{H}_f$ is the Hessian of $f$ evaluated at $\bq=0$. Taking the Legendre transform gives
\beq
g(\bxi) \sim \frac{1}{2} (\bxi - \bxi_*) \cdot \mathsf{H}_f^{-1} \cdot  (\bxi - \bxi_*).
\eeq
In this approximation the concentration is 
\beq
C(\bx,t|\bx_0) \asymp \e^{-(\bx - \bxi_* t) \cdot \mathsf{H}_f^{-1} \cdot  (\bx - \bxi_* t)/(2t)}
\eeq
corresponding to the solution of \eqn{effdiff} with
\beq \lab{effdiff3}
\Pe \av{\bu}=\bxi_* \inter{and} \keff=\mathsf{H}_f/2.
\eeq
This result also follows from \eqn{effdiff2} noting that the mean and covariances that appear on the left-hand sides are given by the first and second derivatives with respect to $\bq$ of the cumulant generating function $\log \E \e^{\bq \cdot \bX} \sim f(\bq) t$ evaluated $\bq=0$. 

Since the diffusive approximation is recovered from the large-deviation results by an expansion for small $\bq$, it can be expected that the method of homogenisation is equivalent to the perturbative solution of the eigenvalue problem \eqn{eig1} or \eqn{eig2}. This is plainly the case. Consider the periodic-flow configuration and assume that $\av{\bu}=0$ for simplicity. Expanding
\beq
\phi = 1 + |\bq| \phi_1 + |\bq|^2 \phi_2 + \cdots \inter{and}
f = |\bq| \alpha_1  + |\bq|^2 \alpha_2  + \cdots, 
\eeq
and introducing this into \eqn{eig1} yields at $O(q)$,
\[
\nabla^2 \phi_1 - \Pe \, \bu \cdot \nabla \phi_1 + \Pe \, \bu \cdot \hat{\bq} = \alpha_1,
\]
where $\hat{\bq}=\bq/|\bq|$ is a unit vector. Averaging this equation gives that $\alpha_1= \Pe \av{\bu \cdot \hat{\bq}}=0$. The solution $\phi_1$ is then written as
\[
\phi_1 = - \hat{\bq} \cdot \bchi
\]
in terms of the periodic, zero-average solution $\bchi$ of the so-called cell problem
\beq \lab{cellprob}
\nabla^2 \bchi - \Pe \, \bu \cdot \nabla \bchi = \Pe \, \bu.
\eeq
\citep[see][\S2.1]{majd-kram}. At order $O(q^2)$, the eigenvalue problem reduces to
\[
\nabla^2 \phi_2 - \Pe \, \bu \cdot \nabla \phi_2 - 2 \hat{\bq} \cdot \nabla \phi_1 + \Pe \, (\bu \cdot \hat{\bq}) \phi_1 = \alpha_2.
\]
Averaging gives
\[
\alpha_2 = 1 + \Pe \langle  (\bu \cdot \hat{\bq}) \phi_1 \rangle = 1 + \hat{\bq}_i  \langle \nabla \bchi_i \cdot \nabla \bchi_j \rangle  \hat{\bq}_j, 
\]
where the second equalities follows after some manipulations using \eqn{cellprob}  \citep[see][ p.\ 251 for details]{majd-kram}. This corresponds to an effective diffusivity with components
\[
\keff_{ij} = \frac{1}{2}\left({H_f}\right)_{ij} =  \delta_{ij} +  \langle \nabla \bchi_i \cdot \nabla \bchi_j \rangle, 
\]
which is the standard homogenisation result. An analogous computation detailed in Appendix \ref{sec:exp} shows how the homogenisation results for shear flows are recovered from  the large-deviation calculation.

The perturbative solution of the eigenvalue problem \eqn{eig1} offers a route for the systematic improvement of the diffusive approximation. Such improvements, which have been derived for shear flows by \citet{chat70,chat72}, \citet{merc-robe} and others \citep[see][for a review]{youn-jone}, extend the diffusion equation \eqn{effdiff} to include higher-order spatial derivatives and increase the accuracy of the approximation for $t \gg 1$. They lead to effective equations of the form
\beq \lab{gendiff}
\partial_t C + \Pe \av{\bu} \nabla \cdot C  = \keff_{ij} \partial_{ij} C +  \keff_{ijk}^{(3)}\partial_{ijk} C + 
 \keff_{ijkl}^{(4)} \partial_{ijlk} C + \cdots,
 \eeq
where summation over repeated indices is understood and we have introduced higher-order effective tensors $\keff_{ijk}^{(3)}$, etc. The behaviour of the large-deviation function $f(\bq)$ as $\bq \to 0$ encodes all these tensors. This can be deduced from the large-deviation form \eqn{largedevi} of the concentration which implies that $\partial_t C \sim f(\bq) C$ and $\nabla C \sim -\bq C$. Combining these formally leads to the effective equation
\beq \lab{efff} 
\partial_t C = f(-\nabla) C.
\eeq
Comparison with \eqn{gendiff} shows that the various effective tensors that appear are given as derivatives of $f(\bq)$ at $\bq=0$. Hence they can be computed by continuing the perturbative solution of the eigenvalue problem \eqn{eig1} to higher orders in $q$. This is demonstrated to $O(q^3)$ for shear flows in Appendix \ref{sec:exp}.

Another kind of improvement captures finite-time effects, specifically the fact that the mean and variance of the particle position have  $O(1)$ corrections to their linear growth which depend on initial conditions. These corrections have been computed for some shear flows \citep{aris56,merc-robe,youn-jone} and termed `initial displacement' and `variance deficit'.
 Although we do not consider them further in what follows, it can noted that Eq.\ \eqn{backkol} for the moment generating function is exact. Its solution for finite time can be expressed as a series of the form $\sum_n A_n(\bq) \exp(f_n(\bq) t) \phi_n^\dagger(\bx)$, where $f_n(q)$ and $\phi_n^\dagger(\bx)$ denote the complete set of eigenvalues and eigenfunctions of \eqn{eig2}. The constants $A_n(\bq)$ can be determined from the initial condition of the concentration. It is clear, then, that the first 2 terms in the Taylor expansion of $A_0(\bq)$, where the $n=0$ mode corresponds to the eigenvalue $f_0(\bq)=f(\bq)$, determine the initial displacement and variance deficit; the other eigenvalues $f_n(\bq),\, n\ge 1$ contribute to exponentially small corrections.

In the rest of the paper, we apply the results of this section to several specific shear and periodic flows. We start with the case of shear flows for which the eigenvalue problems \eqn{eig1} and \eqn{eig2} simplify considerably.

\section{Shear flows} \label{sec:shear}

Consider the advection by a parallel shear flow $\bu=(u(y),0)$ in two dimensions, in a channel of width $2a$ corresponding to $-1 \le y \le 1$ for the dimensionless coordinate $y$. Without loss of generality (exploiting a suitable Galilean transformation as necessary) the velocity can be assumed to satisfy
\beq \lab{zeroav}
\langle u \rangle = \frac{1}{2} \int_{-1}^1 u(y) \, \d y = 0.
\eeq 
Because it is the longitudinal dispersion that is of interest, we modify 
\eqn{largedevi} and take the large-deviation form of the concentration to be
\beq \lab{Cdisp}
C(\bx,t) \sim t^{-1/2} \phi(y,\xi) \e^{-t g(\xi)}, \quad \textrm{where} \ \ \xi = \Pe^{-1} x/t,
\eeq
assuming $\bx_0=0$. Similarly, we write the moment generating function as
\beq \lab{w-disp}
w(q,\bx,t) = \E \e^{\Pe^{-1} q X} \asymp \e^{\Pe^{-1} q x + f(q) t} \phi^\dagger(y).
\eeq
Note that $g$ and $f$ depend only on the longitudinal variables $\xi$ and $q$ and that $\phi$ can be taken $x$-independent because of the $x$-independence of the flow.
The factors $\Pe^{-1}$ are introduced in \eqn{Cdisp}--\eqn{w-disp} for convenience: they lead to a Legendre pair of functions $f(q)$ and $g(\xi)$  that are independent of $\Pe$ in the limit $\Pe \to \infty$, at least for $\xi, \, q=O(1)$. The eigenvalue problem \eqn{eig1} then reduces to the Schr\"odinger form
\beq \lab{eig-disp}
\dt{^2 \phi}{y^2} + \left(q u(y) + \Pe^{-2} q^2 \right) \phi = f(q) \phi. 
\eeq
This one-dimensional eigenvalue problem is completed by the no-flux boundary conditions
\beq \lab{bc-disp}
\dt{\phi}{y}(-1)=\dt{\phi}{y}(1)=0.
\eeq
Note that the operator in \eqn{eig-disp} is self adjoint and hence the same equation arises for the eigenvalue problem \eqn{eig2} for $\phi^\dagger$ associated with the moment generating function. 
Note also that \eqn{eig-disp} can be derived more directly using the Feynman--Kac formula. To see this, write \eqn{sde} explicitly as
\beq \lab{sdeshear}
\d X = \Pe \, u(Y) \d t + \sqrt{2} \d W_1, \quad
\d Y = \sqrt{2} \d W_2,
\eeq
and note that $Y(t)=y + \sqrt{2} W_2$. The generating function \eqn{w-disp} then becomes
\[
w(q,\bx,t)=\E \e^{q \left( \Pe^{-1} (x + \sqrt{2} W_1) + \int_0^t u(y + \sqrt{2} W_2) \, \d t' \right)} = \e^{\Pe^{-1} q x + \Pe^{-2} q^2 t } \E \e^{ q\int_0^t u(y + \sqrt{2} W_2) \, \d t'}.
\]
Using the Feynman--Kac formula \citep[e.g.][]{okse98}, $w$ is seen to satisfy
\[
\partial_t w = \partial_{yy} w + (q u(y) + \Pe^{-2} q^2) w
\]
and hence, for $t \gg 1$, to depend on $t$ as $w \asymp \exp(f(q)t)$ with $f(q)$ the principal eigenvalue in \eqn{eig-disp}. 

Alternatively, \eqn{eig-disp} is  obtained when seeking normal-mode solutions of the form $C(\bx,t) = \phi(k,y) \exp\left(\i (k x - \omega t)\right)$  to the advection--diffusion equation \eqn{ad-dif} provided that the identification $q=\i k$ and $f(q)=-\i \omega(k)$ is made. The large-deviation form of $C$ is then recovered by applying the steepest-descent method to the normal-mode expansion of $C(x,y,t)$. The large-deviation approach makes it clear that the saddle point in the $k$ plane is on the imaginary axis with a purely imaginary associated frequency $\omega=\i f(\i k)$.

Below we solve \eqn{w-disp}--\eqn{bc-disp} numerically for some classical shear flows. Several general remarks can already be made. First, the term proportional to $\Pe^{-2}$ in \eqn{eig-disp} is associated with longitudinal (molecular) diffusion. For $q=O(1)$, it can be neglected for  $\Pe \gg 1$, leading to the simpler eigenvalue problem 
\beq \lab{eig-disp2}
\dt{^2 \phi}{y^2} + q u(y)  \phi = f(q) \phi 
\eeq
which makes clear that $f(q)$ and hence $g(\xi)$ are independent of $\Pe$ in the limit $\Pe \to \infty$ with $q, \, \xi = O(1)$. The large-deviation form of $C(\bx,t)$ can be written in terms of dimensional variables $x_*$ and $t_*$ as
\beq \lab{shearscale}
C(\bx_*,t_*) \asymp \e^{-a^{-2} \kappa t_* g(x_*/(Ut_*))},
\eeq
and its range of validity as $\kappa t_*/a^2 \gg 1$ and $x_* = O(U t_*)$. In what follows, we mostly concentrate on the limit $\Pe \to \infty$ and solve \eqn{eig-disp2} rather than \eqn{eig-disp}: the effect of the neglected longitudinal diffusion on $f(q)$ is straightforward, since it simply adds $\Pe^{-2} q^2$, but the corresponding change in $g(\xi)$ is somewhat more complicated.  
It is nonetheless a simple matter to estimate the size of $q$ for which the neglect of longitudinal diffusivity ceases to be a good approximation.

Second, the perturbative solution of eigenvalue problem \eqn{eig-disp} for $|q| \ll 1$, provides an effective diffusivity as sketched in \S\ref{sec:hom}. In terms of $f(q)$, the dimensional effective diffusivity is expressed from \eqn{shearscale} as
\beq \lab{effdiff4}
\keff_* = \frac{a^2 U^2}{2 \kappa} f''(0),
\eeq
and is inversely proportional to the molecular diffusivity in the limit $\Pe \to \infty$. The perturbative calculation carried out in Appendix  \ref{sec:exp} gives
\beq \lab{f''shear}
\frac{1}{2} f''(0) = \langle \left(\int_{-1}^y u(y') \, \d y' \right)^2 \rangle .
\eeq
and recovers the explicit form of $\keff_*$ as obtained using homogenisation \citep[e.g.][]{majd-kram,cama-et-al}. The first of the corrections to the diffusive approximation of \citet{merc-robe} and \citet{youn-jone} is also computed in Appendix \ref{sec:exp}.

Third, the asymptotics of \eqn{eig-disp2} indicates that $f(q)$ tends to $u_\pm q$ as $q \to \pm \infty$, where $u_\pm$ denote the maximum  and minimum velocities in the channel. This can be seen by noting that $f(q)$ is the lowest eigenvalue of a Schr\"odinger operator which, in the semiclassical limit $|q| \to \infty$, is given by the minimum of the potential $q u(y)$ \citep[e.g.][]{simo83}.
The implication,  as discussed in \S\ref{sec:prob}, is that $g(\xi) \to \infty$ as $\xi \to u_\pm$. Physically, this corresponds to the fact that fluid particles have longitudinal velocities in the range $[u_-,u_+]$;  changes in the concentration therefore propagate at finite speeds and the concentration $C$ is compactly supported for $x_* \in [u_- t_* , u_+ t_*]$. This is only an approximation of course: when longitudinal molecular diffusion is taken into account, there is no limit on the propagation speed. It is readily seen that the term $\Pe^{-2} q^2$ becomes comparable to $u_\pm q$ in $f(q)$ for $q=O(\Pe^2)$ and that the rate function is approximately the diffusive $g(\xi) \sim \Pe^2 (\xi - u_\pm)^2/4$ for $\xi$ near $u_+$ ($u_-$) or larger (smaller). This form of $g$ can also be shown to arise from an application of the \citet{frei-went12} small-noise large-deviation theory and is controlled by a single maximum-likelihood trajectory. (This applies only when $q$ is sufficiently large: the dimensional expression \eqn{shearscale} makes this clear, with an argument of the exponential  that scales like $\kappa$ whereas the small-noise large deviation necessarily leads to a $\kappa^{-1}$ scaling, corresponding to a $\Pe^2$ factor with our non-dimensionalisation.)

Finally, we note that the eigenfunctions $\phi(y,\xi)$, where the $\xi$ dependence is inferred from the $q$-dependence using $\xi=f'(q)$, have a simple interpretation. For $\xi>0$ the amount of scalar at $y$ for $x > \xi t$ can be approximated as
\beq \lab{cond}
 \int_{\xi t}^\infty C(x,y,t) \, \d x \asymp \phi(\xi,y) \e^{-t g(\xi)},
\eeq
since, by the convexity of $g$, the integral is dominated by the  contribution of the endpoint $x=\xi t$. Therefore
$\phi(y,\xi)$ gives the scalar distribution across the shear flow of particles with average speed greater than $\xi >0$. Similarly, for $\xi<0$, $\phi(y,\xi)$ gives the distribution of particles with speed less than $\xi$.

\subsection{Couette flow} \label{sec:couette}

We now examine classical shear flows, starting with the plane Couette flow
\beq \lab{couette}
u(y)=y.
\eeq
The dispersion in this flow is illustrated in Figure \ref{fig:pdfcouette}. The figure shows how the diffusive and large-deviation approximations provide a good approximation in the core of the scalar distribution and how only large deviation captures the tails.
Figure \ref{fig:pdfcouette} does not resolve the tails of $C(x,t)$ with sufficient detail to assess the validity of the large-deviation approximation fully, however. In what follows, we test systematically the large-deviation prediction for $f(q)$, defined as
\beq \lab{f-disp}
f(q) = \lim_{t \to \infty} \frac{1}{t} \log \E \e^{\Pe^{-1} q X(t)}
\eeq
with our shear-flow scaling, by comparing the value obtained by solving the eigenvalue problem \eqn{eig-disp} for a range of $q$ with careful Monte Carlo estimates.
The eigenvalue problem is solved using a finite-difference scheme. (An exact solution can be written in terms of Airy functions, but it is not particularly illuminating). The Monte Carlo estimates approximate the right-hand side of \eqn{f-disp} as an average over a large number of solutions of \eqn{sdeshear}. However, a straightforward implementation does not provide a reliable estimate for $f(q)$ except for small values of $q$. This is because $f(q)$ for moderate to large $q$ is controlled by rare realisations which are not  sampled satisfactorily. To remedy this, it is essential to use an importance-sampling technique which concentrates the computational effort on these realisations. For the results reported in this paper, we have implemented  a version of Grassberger's \citeyearpar{gras97} pruning-and-cloning technique which we describe in Appendix \ref{sec:resamp}. 

\begin{figure}
\begin{tabular}{cccc}
Couette  \hspace{-0.75cm} & Plane Poiseuille  \hspace{-0.75cm} & Pipe Poiseuille \\
\includegraphics[height=3.8cm]{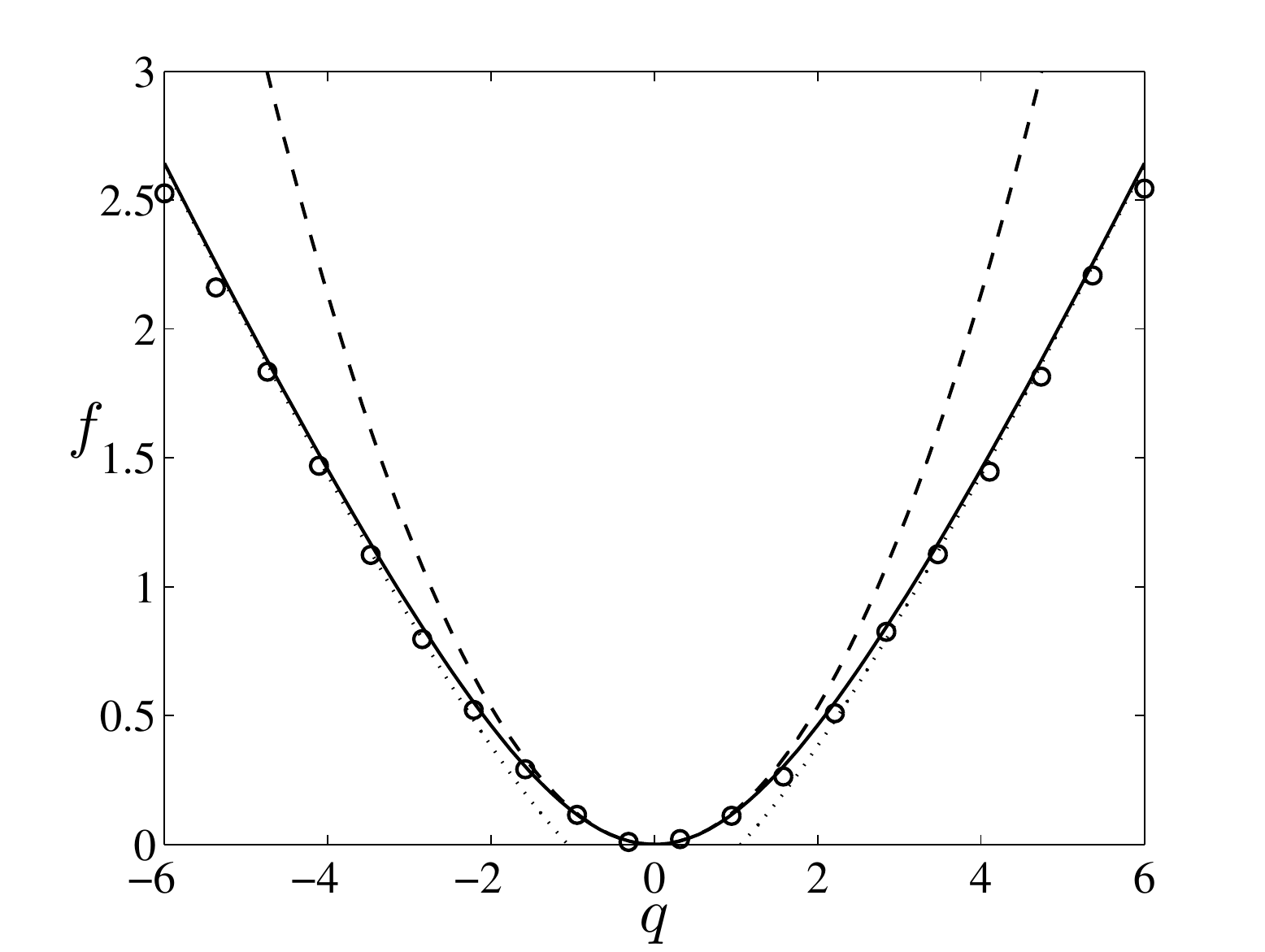} \hspace{-0.75cm} &
\includegraphics[height=3.8cm]{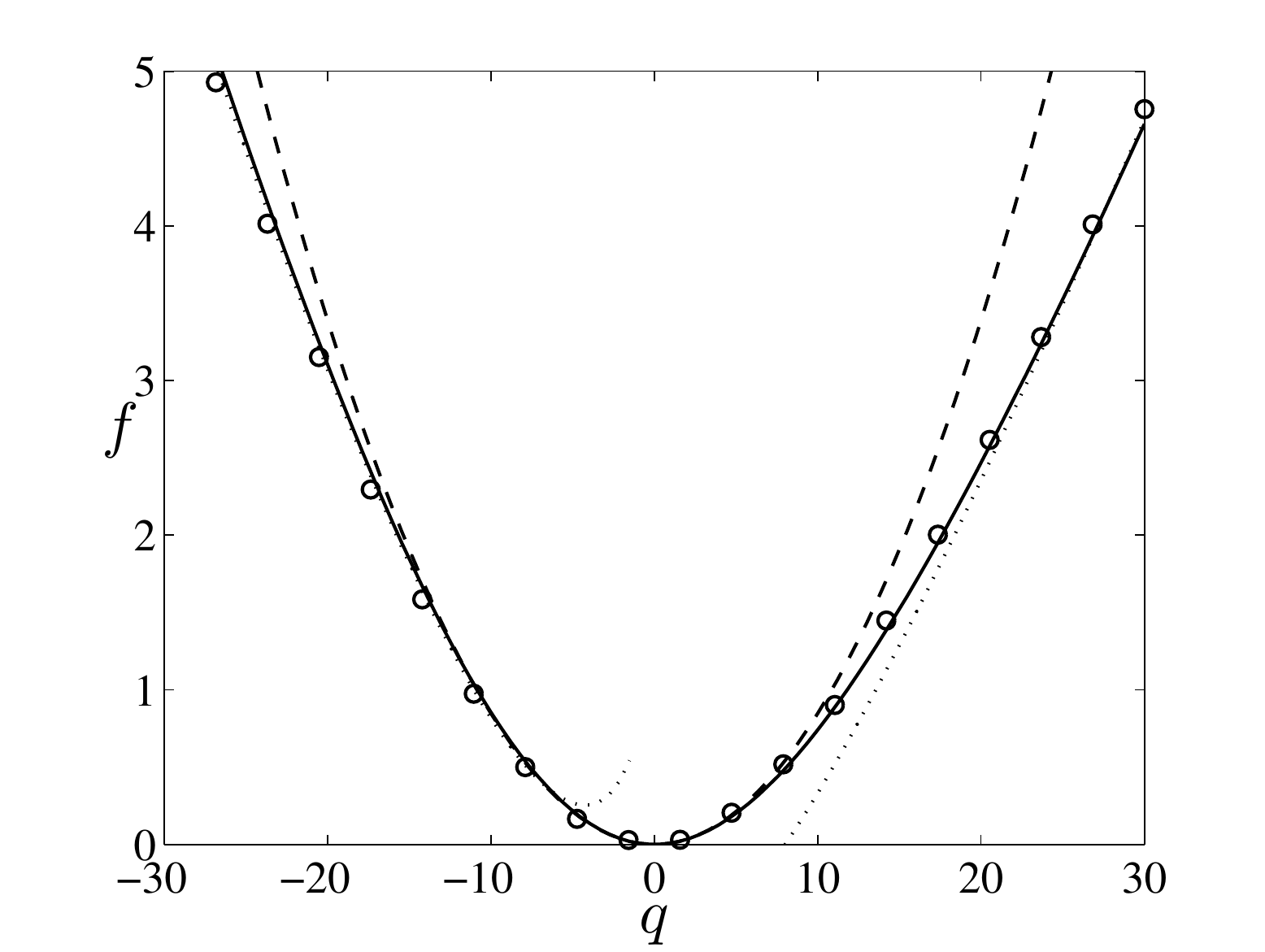} \hspace{-0.75cm} & 
\includegraphics[height=3.8cm]{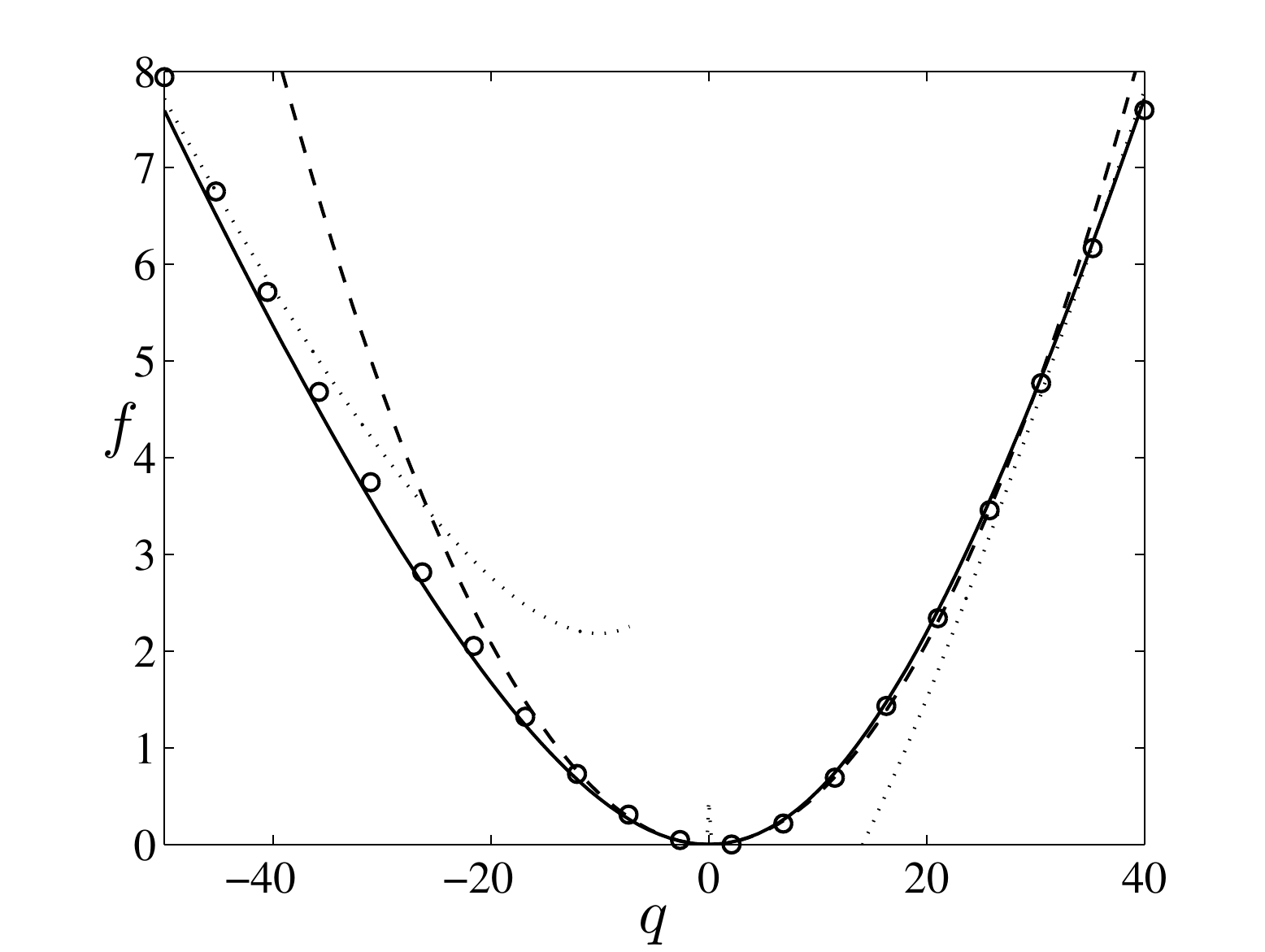} \\
\includegraphics[height=3.8cm]{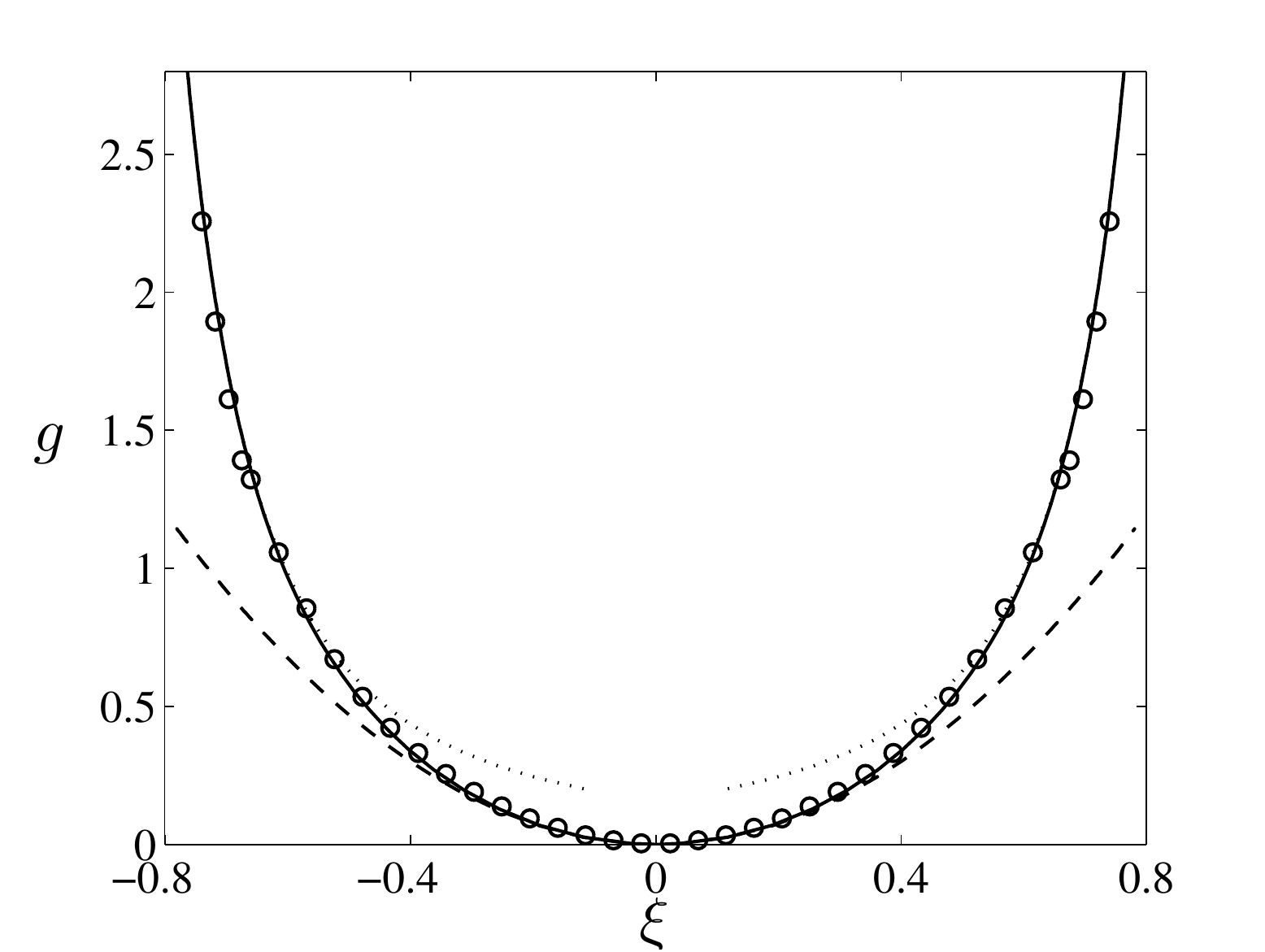}\hspace{-0.75cm} &
\includegraphics[height=3.8cm]{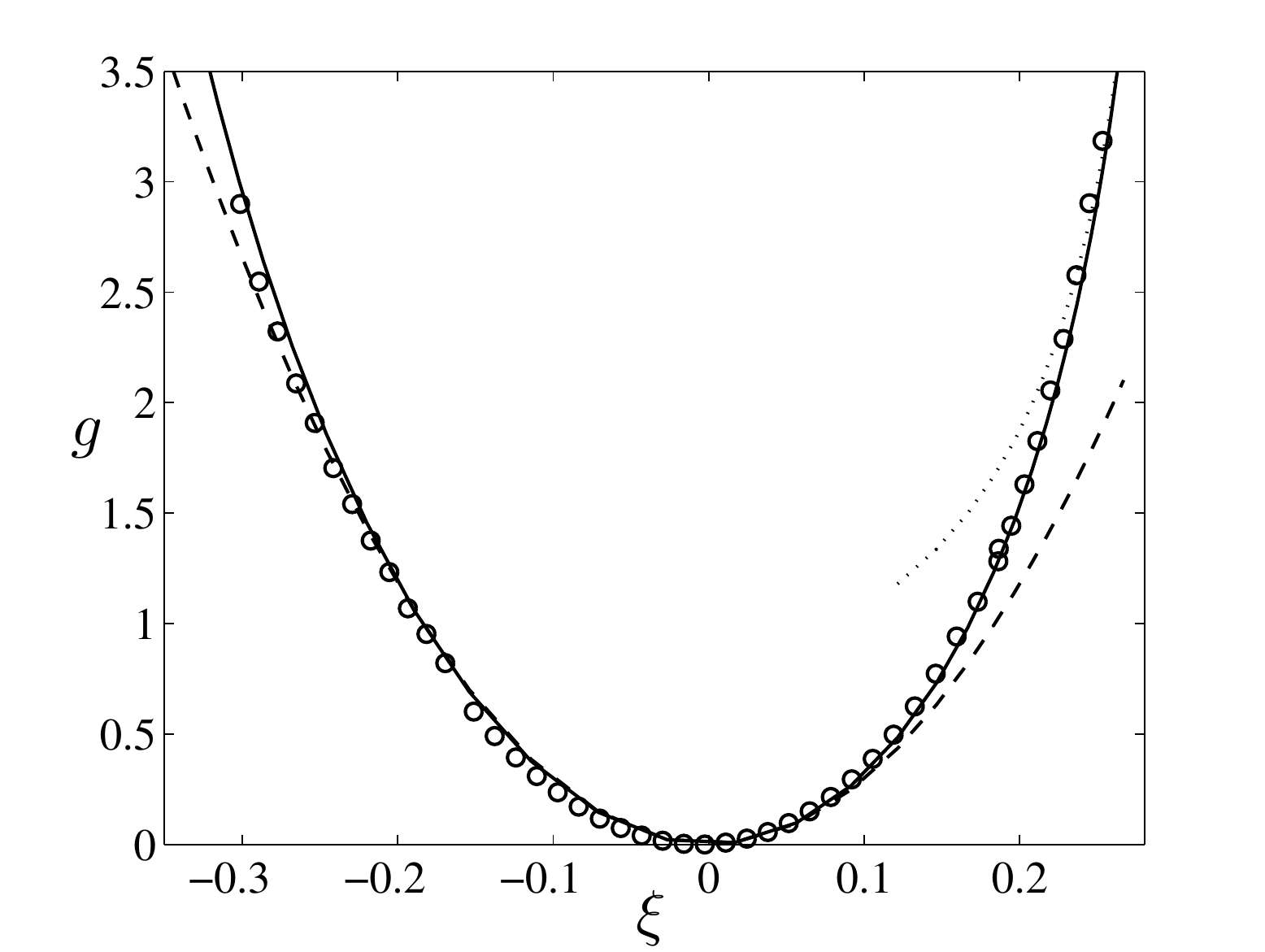} \hspace{-0.75cm} &
\includegraphics[height=3.8cm]{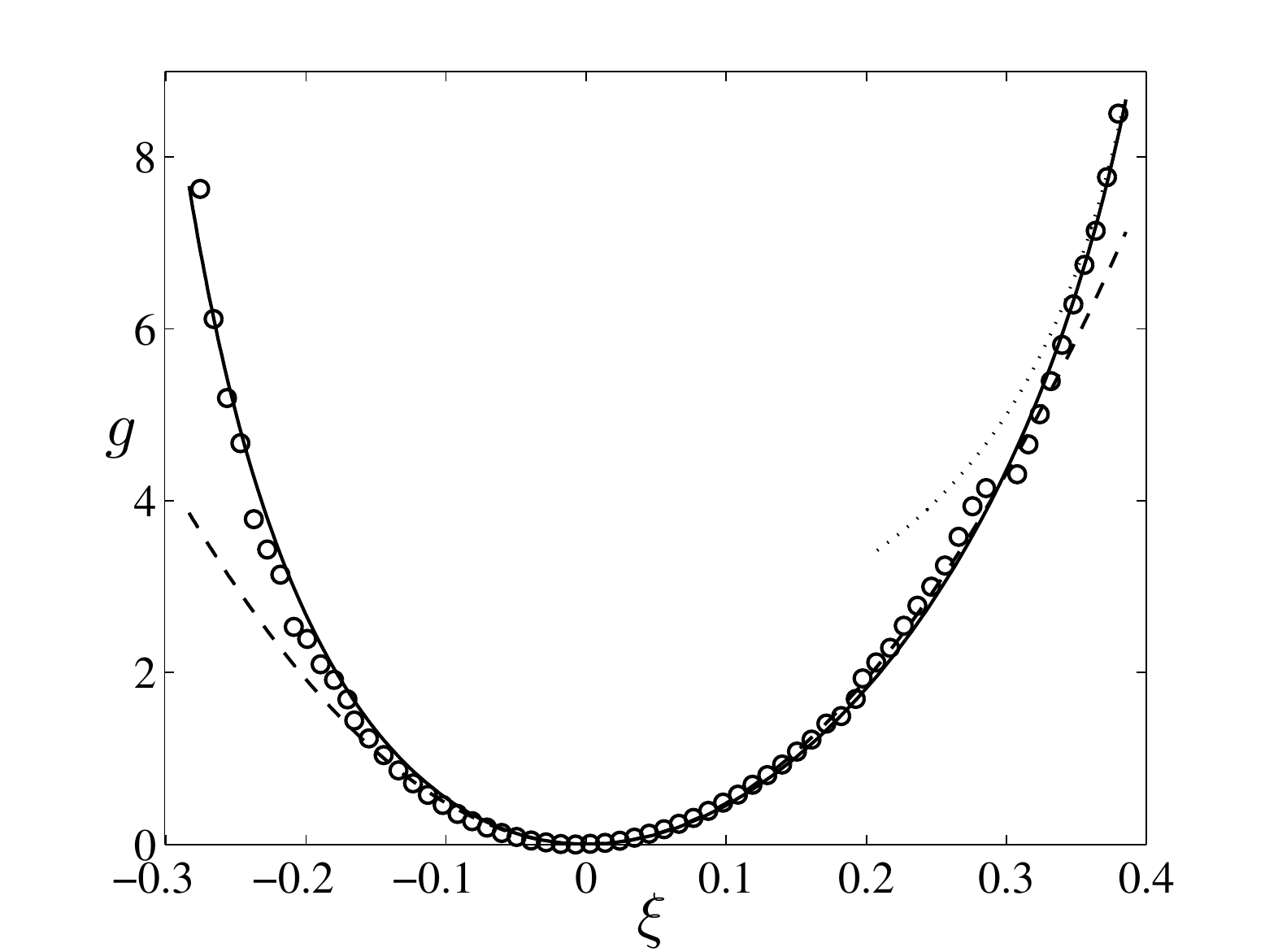} \\
\includegraphics[height=3.8cm]{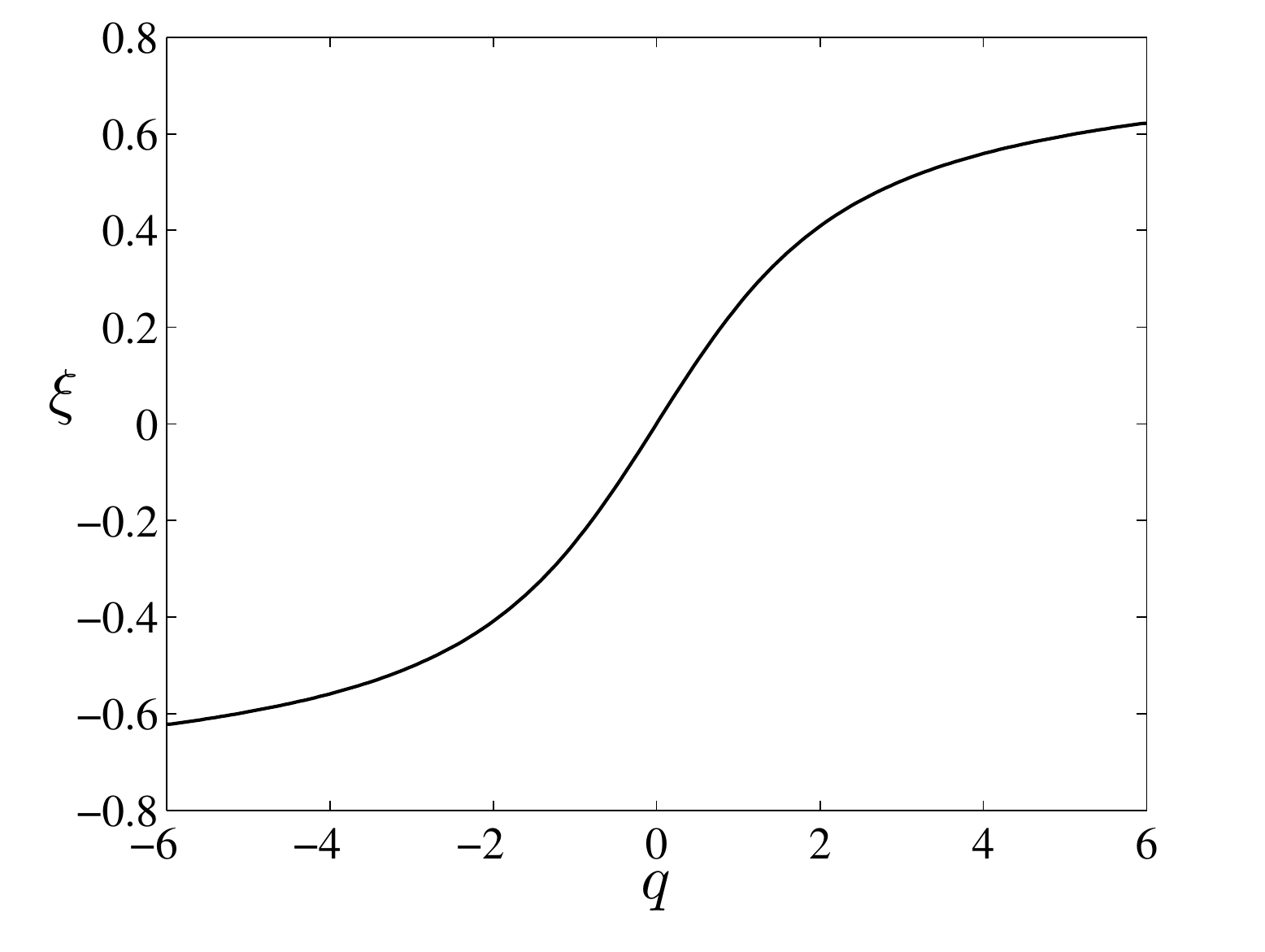} \hspace{-0.75cm} &
\includegraphics[height=3.8cm]{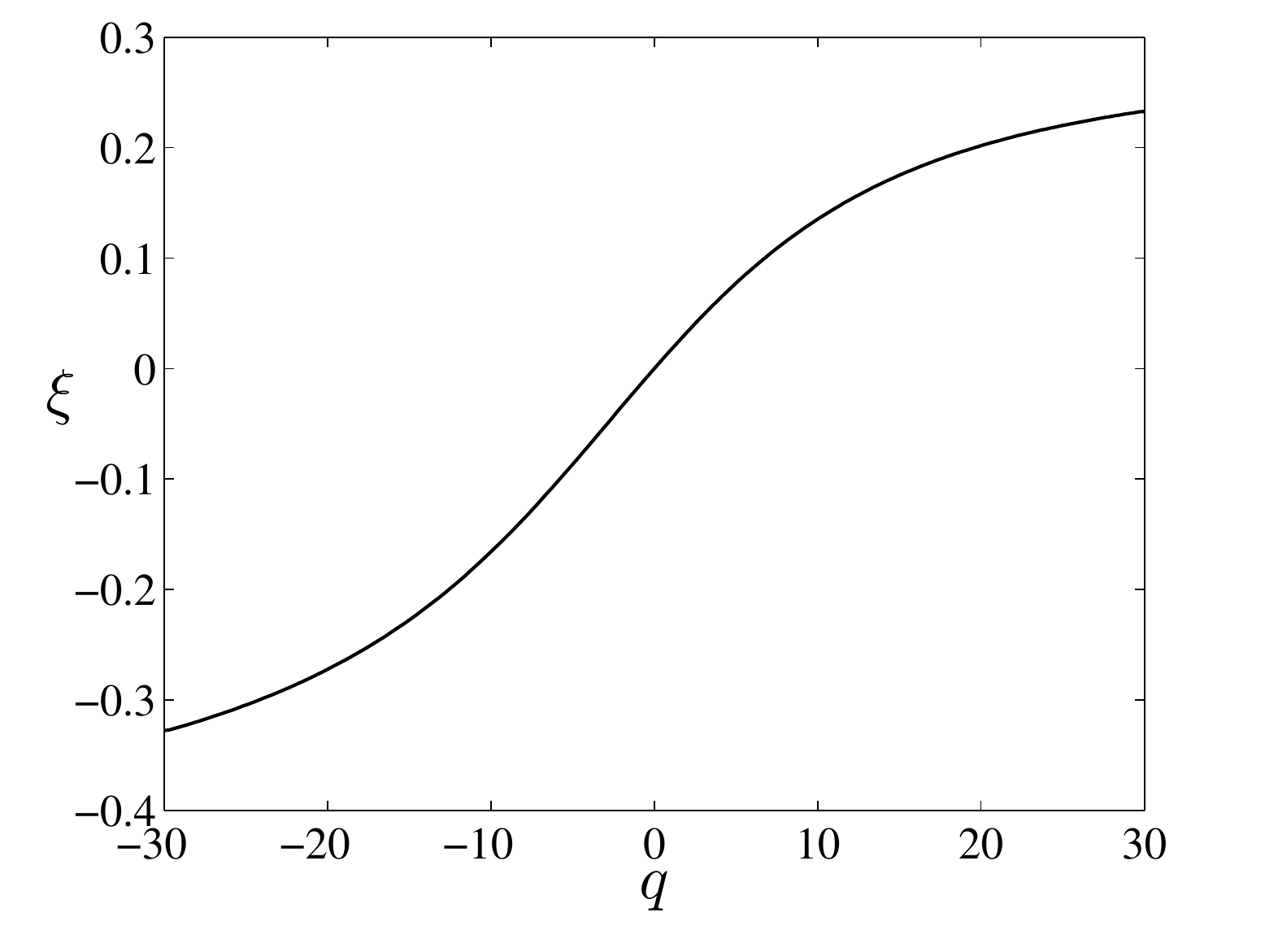} \hspace{-0.75cm} & 
\includegraphics[height=3.8cm]{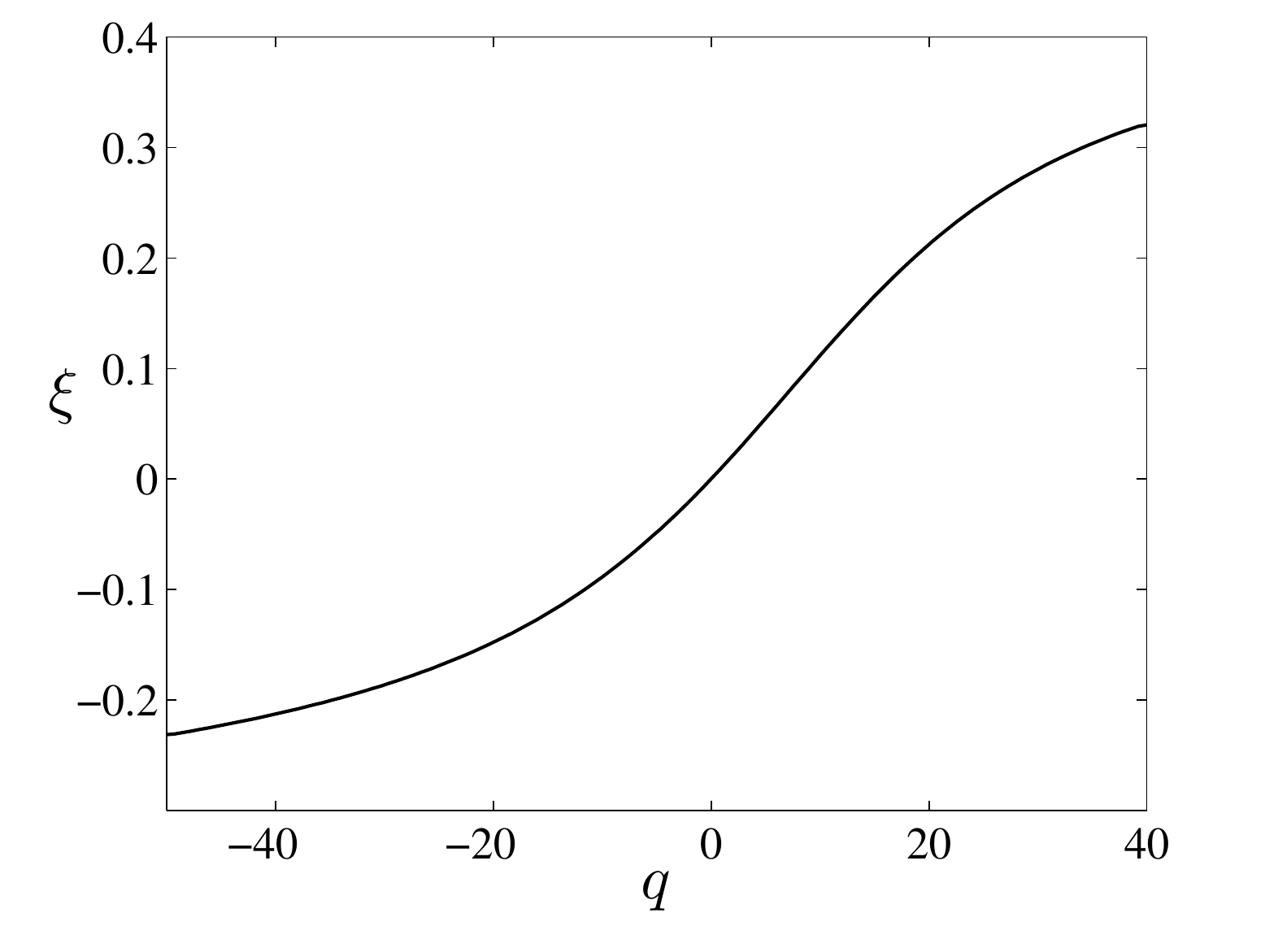} \vspace{-.2cm} \\ 
\includegraphics[height=3.8cm]{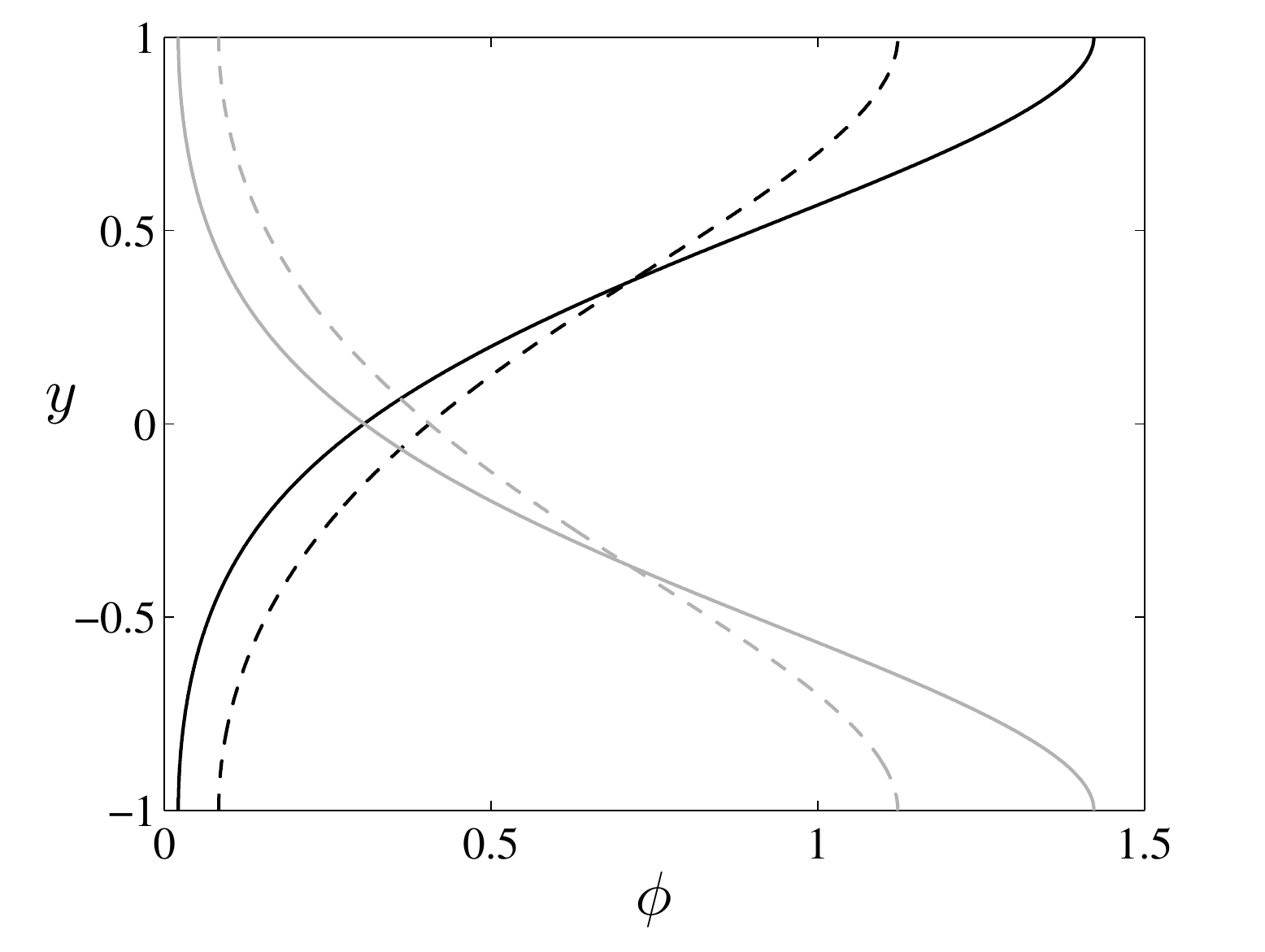} \hspace{-0.75cm} &
\includegraphics[height=3.8cm]{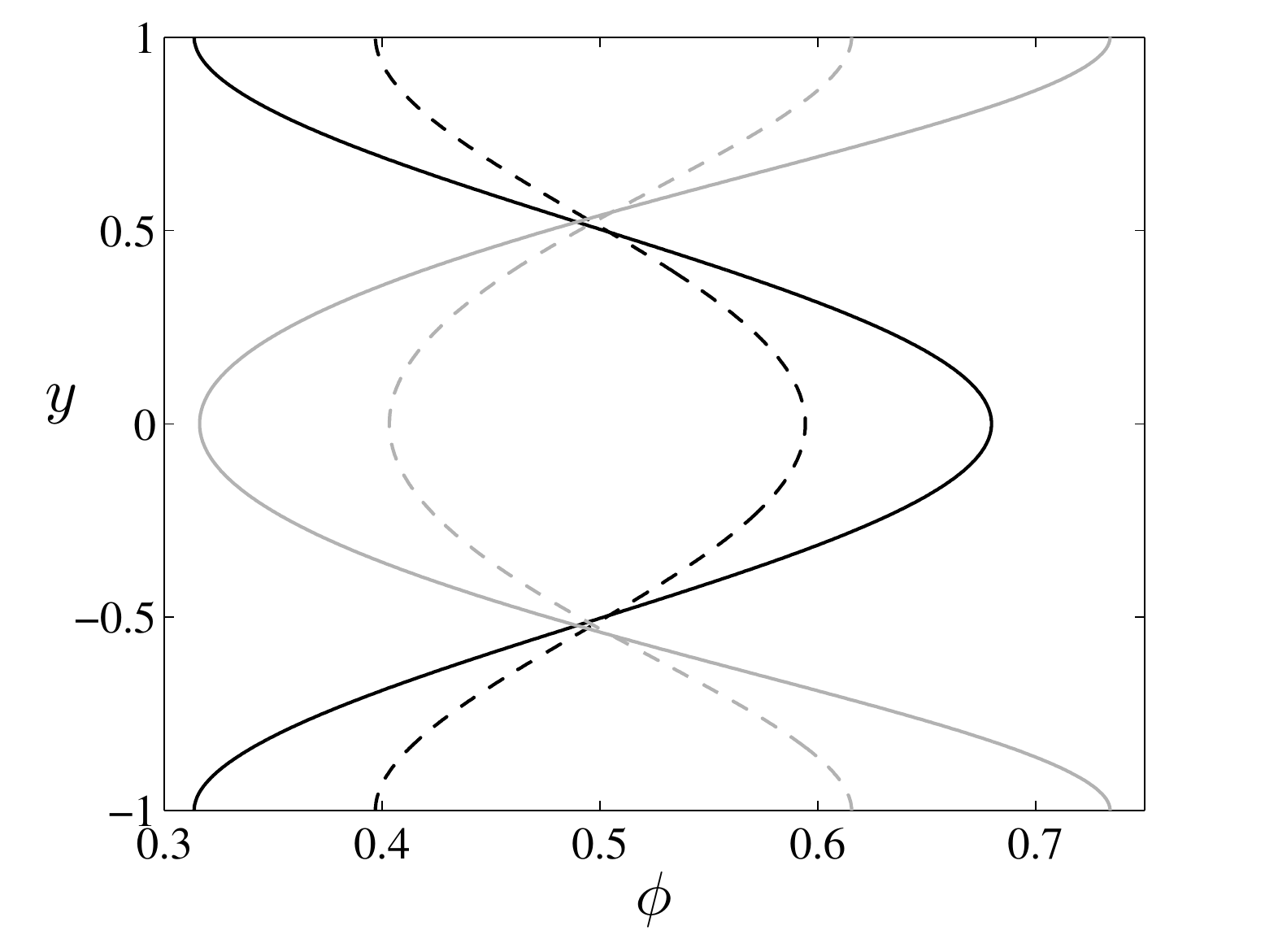} \hspace{-0.75cm} & 
\includegraphics[height=3.8cm]{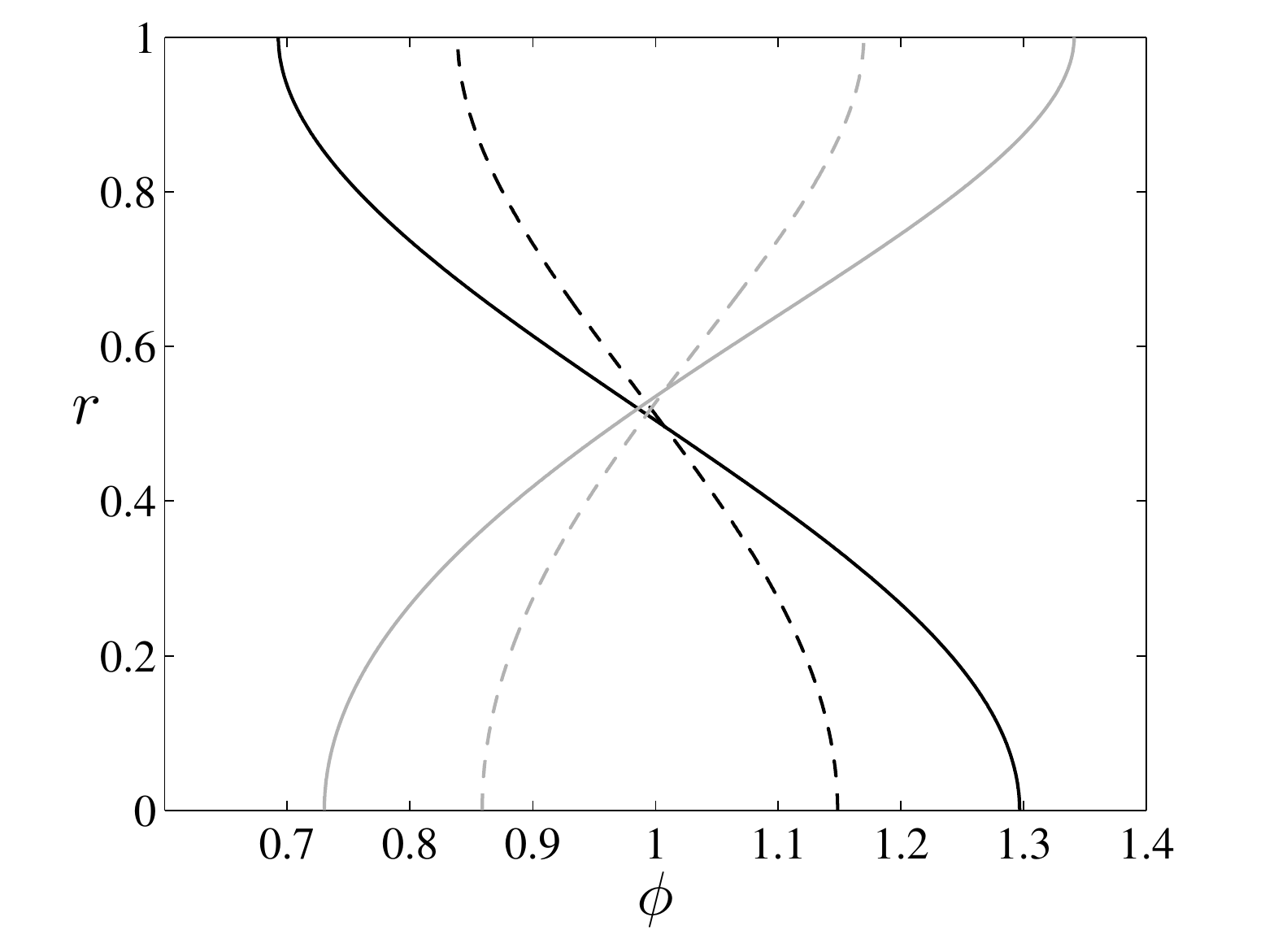} \\
\end{tabular}
\caption{Large-deviation results for Couette, plane Poiseuille and pipe Poiseuille flows. First row: the eigenvalue $f(q)$ obtained by numerical solution of the eigenvalue problem (solid line) is compared with Monte Carlo estimates (symbols). The small-$q$ (diffusive) and large-$q$ asymptotic approximations are also shown (dashed and dotted lines). Second row: the rate function $g(\xi)$ obtained by Legendre transform of the eigenvalue problem solution $f(q)$ (solid line) is compared with direct Monte Carlo estimates (symbols). The asymptotic approximations for small $\xi$  and for $\xi \to u_\pm$, the maximum and minimum flow speeds, are also shown (dashed and dotted lines). (For the two Poiseuille flows, the approximations for $\xi \to u_-$ are not shown because the range of $\xi$ does not extend to their regions of validity.) Third row:  map between $q$ and $\xi=x/t$ derived from the numerical estimate of $f(q)$. Fourth row: eigenfunctions $\phi$ for $q=5,\, 10$ (dashed and solid black lines) and for $q=-5,-10$ (dashed and solid grey lines).} \label{fig:shearall}
\end{figure}

Results for the plane Couette flow are displayed in the leftmost panels of Figure \ref{fig:shearall}.
The top panel shows the eigenvalue and Monte Carlo approximations of $f(q)$ along with asymptotic approximations valid for small and large $q$. 
%
%
%
The small-$q$ approximation for $f(q)$ is found from \eqn{f''shear} as
\beq \lab{smallqcouette}
f(q) \sim   \frac{2}{15} q^2 \quad \textrm{as} \ \ q \to 0.
\eeq
The large-$|q|$ approximation is obtained by noting that for $q \to \pm \infty$, the solution to \eqn{eig-disp2} is localised in boundary layers near $y=\pm 1$. Concentrating on $q \to \infty$, we introduce $
y = 1 - q^{-1/3} Y$ and  $f(q) = q + q^{2/3} \mu$ into \eqn{eig-disp2}. To leading order, this gives
\beq
\dt{^2 \phi}{Y^2} - Y \phi = \mu \phi,
\eeq
with solution $\phi = \Ai(Y+\mu)$ decaying as $Y \to \infty$. Imposing the boundary condition at $Y=0$ gives the equation $\Ai'(\mu)=0$ for $\mu$. Hence we have
\beq \lab{largeqcouette}
f(q) \sim |q| - 1.019 |q|^{2/3} \quad \textrm{as} \ \ |q| \to \infty,
\eeq
using symmetry to deal with $q \to - \infty$. 

The top left panel of Figure \ref{fig:shearall} confirms the validity of the eigenvalue calculation and of the asymptotic estimates. In the case of the $|q|\gg 1$ estimates, a constant is added to \eqn{largeqcouette} to ensure a good match; with this $o(1)$ correction, the asymptotic formula appears to be accurate for $|q|$ as small as $3$, say. The dispersive approximation corresponding to the parabola  \eqn{smallqcouette} overestimates $f(q)$ for all $q$, indicating that this approximation overestimates the speed of dispersion or equivalently the magnitude of the tails of the distribution. 

The rate function $g(\xi)$ is shown in the second row of Figure \ref{fig:shearall}. The solid curve is obtained by Legendre transforming the function $f(q)$ computed by numerical solution of the eigenvalue problem. This is compared with direct Monte Carlo estimates. Again, it is crucial to use importance sampling to obtain a reliable estimate of $g(\xi)$ for $\xi$ not small. We have chosen to integrate a modified dynamics in which particles, instead of simply diffusing in the $y$-direction, also experience of drift towards the wall at $y=1$ (or $y=-1$). A better sampling is obtained because the wall regions control $g(\xi)$ for large  $|q|$; the method is described in Appendix \ref{sec:girs}. The Figure also shows the asymptotic approximations for $g(\xi)$ deduced from \eqn{smallqcouette} and \eqn{largeqcouette} by Legendre transform and  given by 
\beq \lab{asxicouette}
g(\xi) \sim \frac{15}{8} \xi^2 \ \ \textrm{as} \ \ \xi \to 0 \inter{and}
g(\xi) \sim \frac{4 \cdot 1.019^3}{27(1\mp\xi)^2} \ \ \textrm{as} \ \ \xi \to \pm 1.
\eeq 

The match between the values of $g(\xi)$ derived from the eigenvalue problem and those obtained by Monte Carlo sampling provides a direct check on the validity of the large-deviation theory. The discrepancy between the exact $g(\xi)$ and its diffusive approximation confirms that diffusion overestimates the dispersion speed, as inferred already from the plot of $f(q)$. The finite support of the concentration distribution for $\xi \in [-1,1]$, arising from the neglect of longitudinal molecular diffusion, is also hinted at by the large slopes of $g$ for $\xi \approx \pm 0.8$. The large-$|\xi|$ approximation to $g(\xi)$ (with $o(1)$ term fixed by inspection) is seen to be accurate for $|\xi|\ge 0.5$ and could be combined with the small $\xi$ approximation to provide a satisfactory uniform approximation. 

The third panel on the left of Figure \ref{fig:shearall} shows the map between $\xi=f'(q)$ that arises as part of the Legendre transform. This identifies the location $x=\xi t$ which control the corresponding exponential moment $\E \exp(q X)$ for large $t$. Finally, the fourth panel shows profiles of the eigenfunctions $\phi(\xi,y)$ of \eqn{eig-disp} for several values of $q$. According to \eqn{cond}, these give the structure of the concentration profile for $x/t$ larger than $\xi=f'(q)$. Thus, for instance, the eigenfunction for $q=5$ approximately corresponds  to $x/t \ge 0.5$ (see third panel). 
As $q$ and hence $\xi$ increase (or decrease) the profile becomes more and more localised in the region of maximum (or minimum) velocity, that is, near $y=1$ ($y=-1$).  
The eigenfunctions for finite $q$ are to be contrasted with the standard (homogenisation) results on Taylor dispersion which correspond to eigenfunctions that are small, $O(q)$ perturbations to the uniform eigenfunction $\phi=1$.

\subsection{Plane Poiseuille flow} \label{sec:poiseuille}

We next examine the plane Poiseuille flow
\beq \lab{ppoiseuille}
u(y) = 1/3 - y^2.
\eeq
The small-$q$ approximation in this case is
readily found from \eqn{f''shear} to be
\beq \lab{smallqpoiseuille}
f(q) \sim   \frac{8}{945} q^2 \quad \textrm{as} \ \ q \to 0.
\eeq
For $q \gg 1$, the solution is localised around the maximum of the velocity at $y=0$. For the required boundary-layer analysis, we let $y = q^{-1/4} Y$ and $f(q)=q/3+\mu q^{1/2}$ and obtain
\beq
\dt{^2 \phi}{Y^2} - Y^2 \phi = \mu \phi.
\eeq
The solution corresponding to the largest eigenvalue $\mu$ is the Gaussian $v=\exp(-Y^2/2)$, leading to $\mu=- 1$ and
\beq \lab{largeqpoiseuille}
f(q) \sim q/3 - q^{1/2} \quad \textrm{as} \ \ q \to \infty.
\eeq
For $q \ll -1$, the asymptotic treatment is similar to that of the Couette flow: we let $y=1-|q|^{1/3}Y$ and $f(q)=2|q|/3+\mu |q|^{2/3}$ and find that $\phi \sim \Ai(2^{1/3}(Y+\mu/2))$ and hence $\Ai'(2^{-2/3} \mu)=0$. This gives the approximation
\beq \lab{large-qpoiseuille}
f(q) \sim -2q/3 - 1.617 q^{2/3} \quad \textrm{as} \ \ q \to -\infty.
\eeq



The corresponding rate function $g(\xi)$ is derived  by  Legendre transform, yielding the asymptotic behaviours
\beq \lab{smallxipoiseuille}
g(\xi) \sim \frac{945}{32} \xi^2 \ \ \textrm{as} \ \ \xi \to 0, 
\eeq
\beq \lab{large-xipoiseuille}
g(\xi) \sim \frac{1}{4(1/3-\xi)} \ \ \textrm{as} \ \ \xi \to 1/3, \ \ \textrm{and} \ \ g(\xi) \sim \frac{4 \cdot 1.617^3}{27(2/3+\xi)^2} \ \ \textrm{as} \ \ \xi \to -2/3.
\eeq

The numerical and asymptotic results obtained for the plane Poiseuille flow are displayed in the second column of Figure \ref{fig:shearall}. As for the Couette flow, the diffusive approximation \eqn{smallqpoiseuille} and \eqn{smallxipoiseuille} is seen to overestimate the speed of dispersion, leading to an overestimate of $f(q)$ and an underestimate of $g(\xi)$.  The  concentration distribution for the Poiseuille flow is skewed, with $g(\xi)$ increasing faster for $\xi>0$ than $\xi<0$ corresponding to smaller concentrations for $\xi>0$ than for $\xi<0$. The eigenfunctions shown in the bottom panel  illustrate how $f(q)$ for large $q$ (small $q$) and hence $g(\xi)$ for large $\xi$ (small $\xi$) are controlled by motion near the centre (periphery) of the flow. This culminates in the limits $q,\, \xi \to \infty$ ($-\infty$) as the boundary-layer form of the eigenfunctions derived above indicates.



\subsection{Pipe Poiseuille flow} \label{sec:pipe}

We conclude this section by considering the Poiseuille flow  in a pipe, with velocity
\beq \lab{pipe}
u(r) = 1/2 - r^2,
\eeq
where $r=\sqrt{y^2+z^2}$. This flow is three-dimensional, with particles diffusing across the flow in both the $y$- and $z$-directions. While the eigenfunctions for axisymmetric flows $\phi$ can in principle depend on $y$ and $z$ independently, the principal eigenvalue determining $f(q)$ is obtained for axisymmetric $\phi$: $\phi=\phi(r)$. Correspondingly, the eigenvalue problem \eqn{eig-disp2} of plane shear flows is replaced by
\beq \lab{eig-axi}
\frac{1}{r} \dt{}{r} \left( r \dt{\phi}{r} \right) + q u(r) \phi =  f(q) \phi
\eeq
with boundary conditions $\d \phi / \d r = 0$ at $r=0,\,1$.

The small-$q$, diffusive approximation $f(q) \sim \alpha_2 q^2$ for general axisymmetric shear flows is quoted in Appendix \ref{sec:exp} as \eqn{alpha2ax}. For the Poiseuille flow, this gives
\beq \lab{smallqpipe}
f(q) \sim \frac{1}{192} q^2 \quad \textrm{as} \ \ q \to 0.
\eeq
For $q \gg 1$, an approximation to $f(q)$ is derived from \eqn{eig-axi} using a boundary-layer approach: we let $r=q^{-1/4} R$ and $f(q)=q/2+\mu q^{1/2}$ to find the leading-order equation
\beq
\frac{1}{R} \dt{}{R} \left( R \dt{\phi}{R} \right) - R^2 \phi = \mu \phi, 
\eeq
with solution $\phi = \exp(-R^2/2)$, corresponding to $\mu=-2$. Therefore,
\beq \lab{largeqpipe}
f(q) \sim q/2 - 2 q^{1/2} \quad \textrm{as} \ \ q \to \infty.
\eeq
The analysis for $q \ll -1$ is almost identical to that carried out for the plane Poiseuille flow and leads to
\beq \lab{large-qpipe}
f(q) \sim -q/2 - 1.617 q^{2/3} \quad \textrm{as} \ \ q \to -\infty.
\eeq
Computing the Legendre transform of \eqn{smallqpipe}, \eqn{largeqpipe} and \eqn{large-qpipe} yields the corresponding asymptotics results for the rate function, namely
\beq \lab{smallxipipe}
g(\xi) \sim 48 \xi^2 \ \ \textrm{as} \ \ \xi \to 0, 
\eeq
\beq \lab{large-xipipe}
g(\xi) \sim \frac{1}{(1/2-\xi)} \ \ \textrm{as} \ \ \xi \to 1/2, \ \ \textrm{and} \ \ g(\xi) \sim \frac{4 \cdot 1.617^3}{27(1/2+\xi)^2} \ \ \textrm{as} \ \ \xi \to -1/2.
\eeq
Note that \eqn{smallxipipe} recover's Taylor's original result \citep{tayl53}.

The numerical and asymptotic results for the pipe Poiseuille flow are shown in the rightmost panels of Figure \ref{fig:shearall}. The diffusive approximation is seen to mostly overestimate the dispersion speed, although it turns out to be remarkably accurate for $q, \, \xi > 0$. Close inspection shows in fact that there is a range of values of $q, \, \xi>0$ for which diffusion underestimates somewhat the concentration, in contrast to the other cases considered. Note that the skewness for the pipe Poiseuille flow is opposite to that of the plane Poiseuille flow, with larger concentrations predicted for $\xi>0$ than $\xi<0$.

\section{Periodic flows} \label{sec:cell}

We now turn to two-dimensional periodic  flows. The formalism of \S\,\ref{sec:formulation} applies directly: $f(\bq)$ is obtained by solving the eigenvalue problem \eqn{eig1} with periodic boundary conditions for $\phi$. Eq.\ \eqn{eig1} can also be obtained in an alternative manner: because  the advection--diffusion equation \eqn{ad-dif} has periodic coefficients, its solutions can be sought in the Floquet--Bloch form $C(\bx,t)=\phi(\bk,\bx) \exp\left(\i (\bk \cdot \bx-\omega t)\right)$, which leads to  \eqn{eig1} with $\i \bk = \bq$ and $\omega(\bk) = \i f(\bq)$ \citep{bens-et-al, papa95}. This approach gives a representation of the concentration as an integral over $\bk$ whose large-$t$ asymptotics, derived using the steepest-descent method, reduces to the large-deviation form \eqn{largedevi}.

%

\begin{figure}
\begin{center}
\includegraphics[height=6cm]{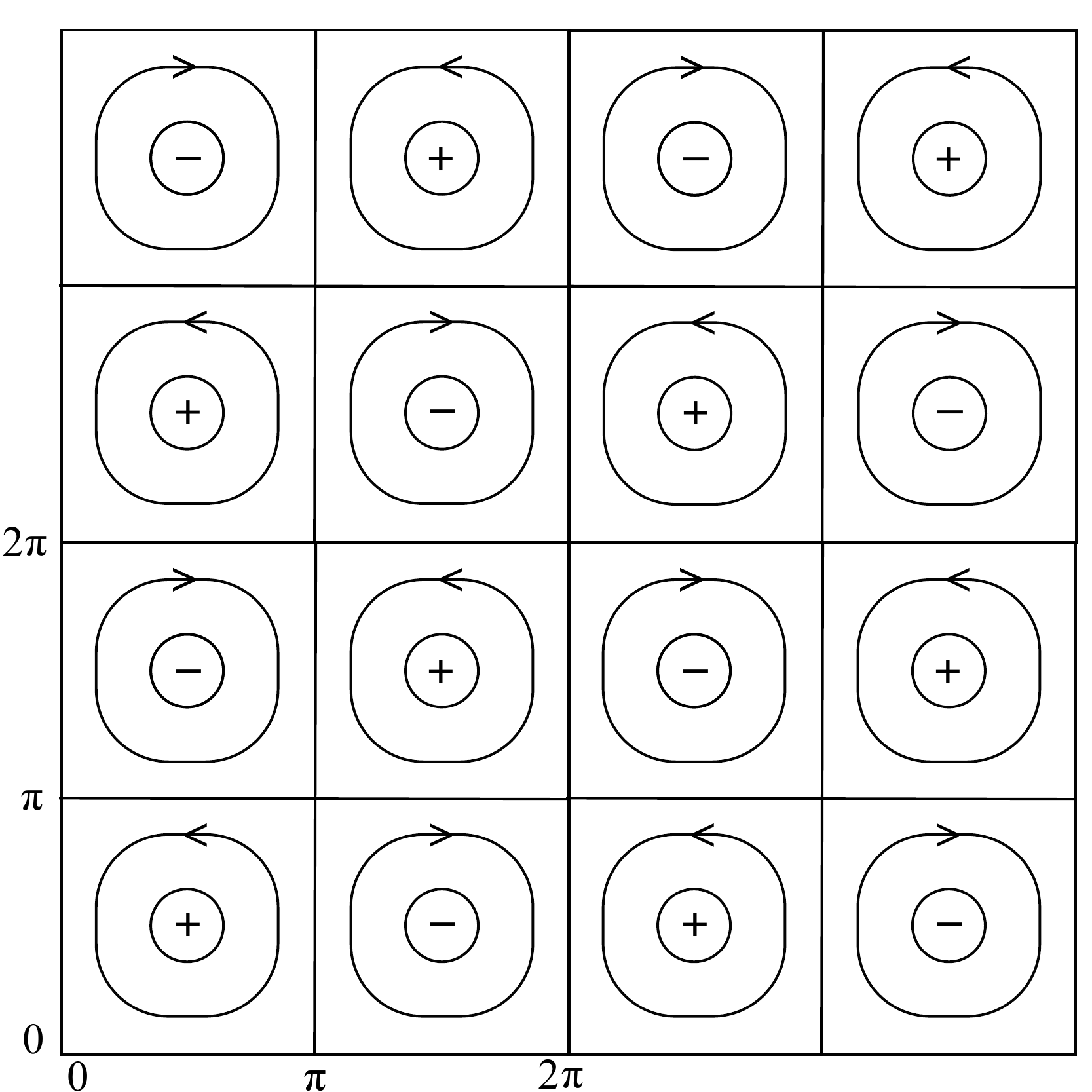}
\caption{Streamlines of the cellular flow \eqn{cellflow1}. Four of the periodic cells are shown.}
\label{fig:cellpicturelarge}
\end{center}
\end{figure}

\begin{figure}
\begin{center}
\includegraphics[height=5cm]{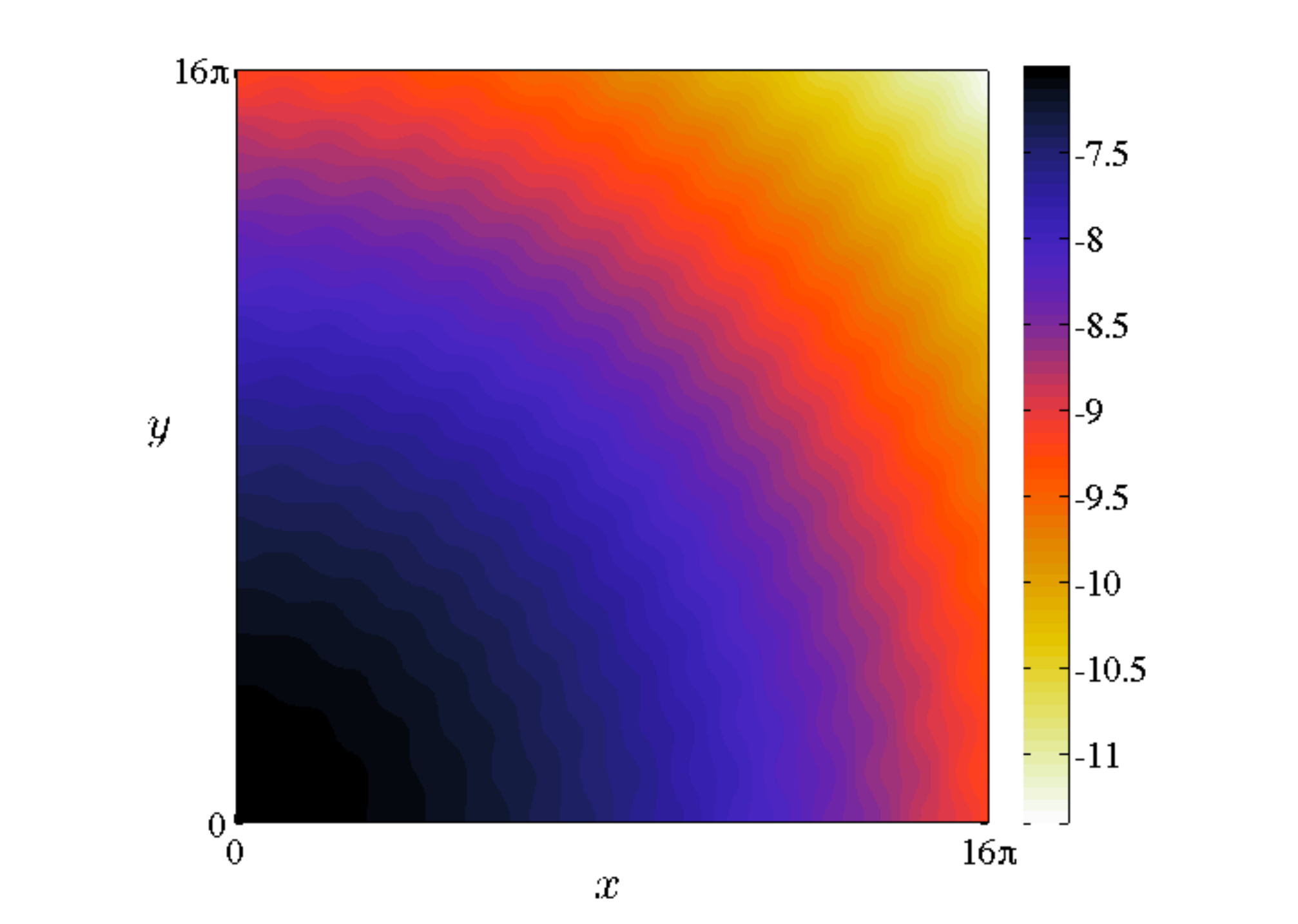} \hspace{-1cm}
\includegraphics[height=5cm]{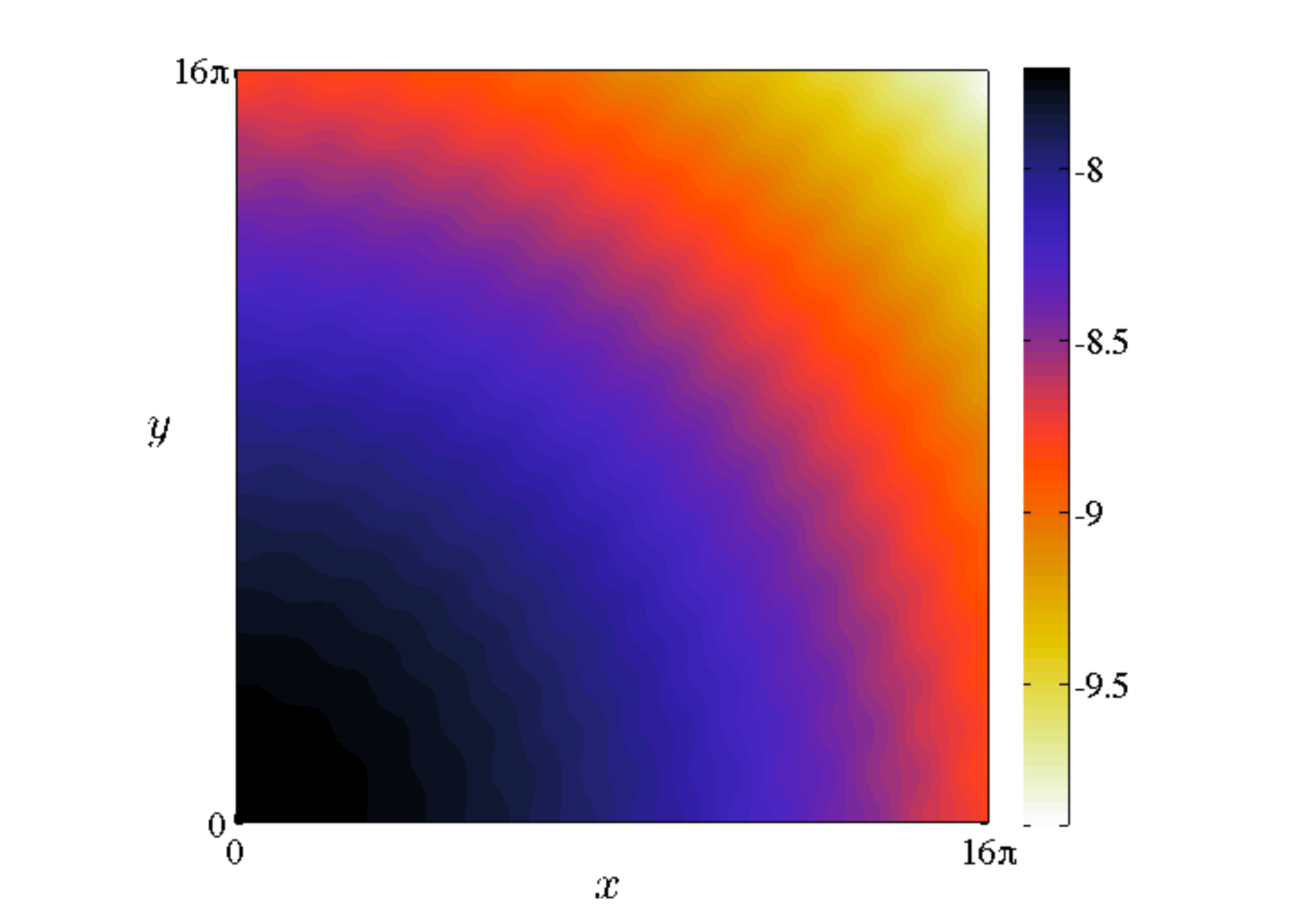}
\caption{(Colour online.) Concentration (in logarithmic scale)  at times $t=250$ (left) and $t=500$ (right) of a scalar initially  released in the central cell of a cellular flow with $\Pe=1$.}
\label{fig:simu1}
\end{center}
\end{figure}

\begin{figure}
\begin{center}
\includegraphics[height=5cm]{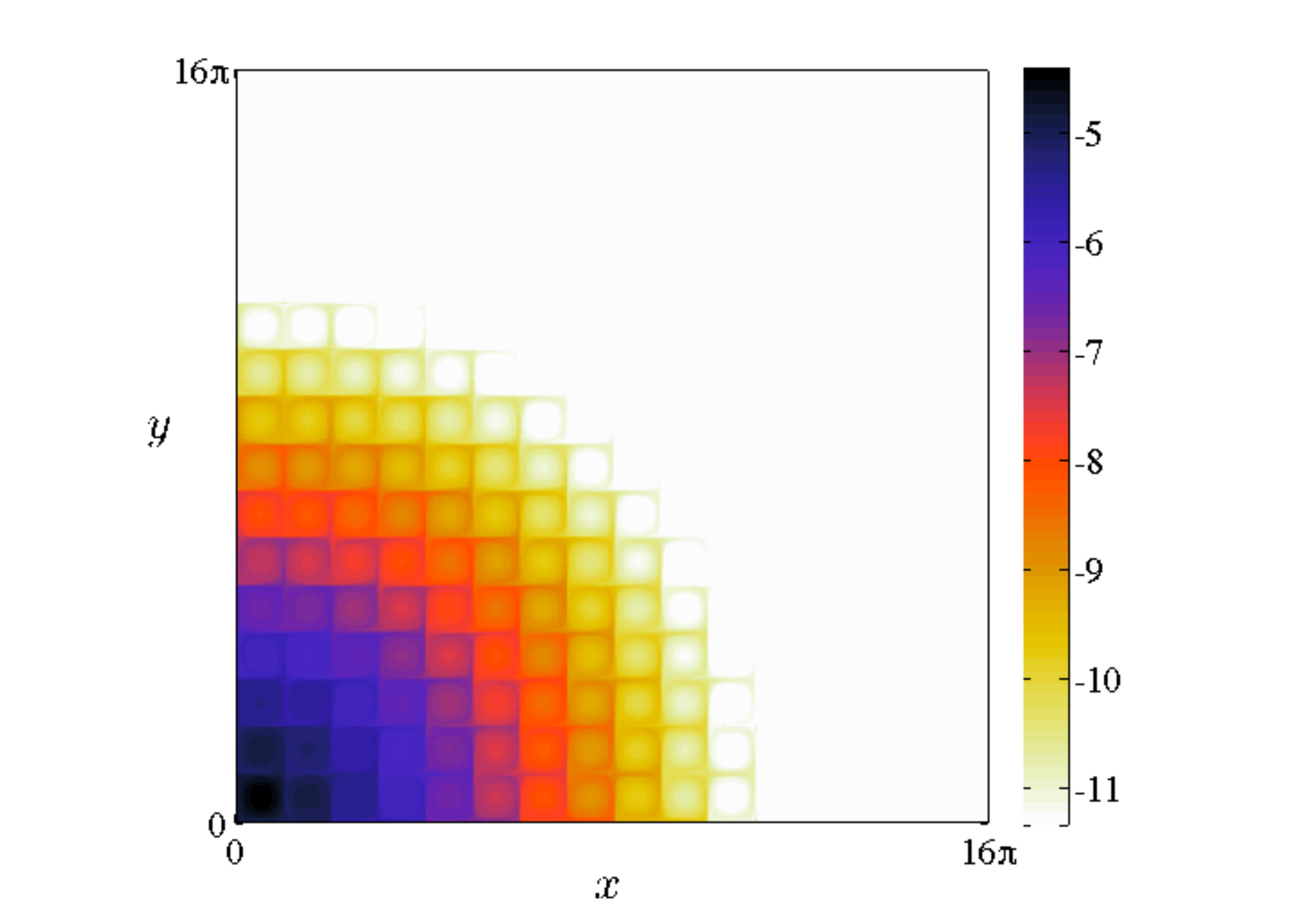} \hspace{-1cm}
\includegraphics[height=5cm]{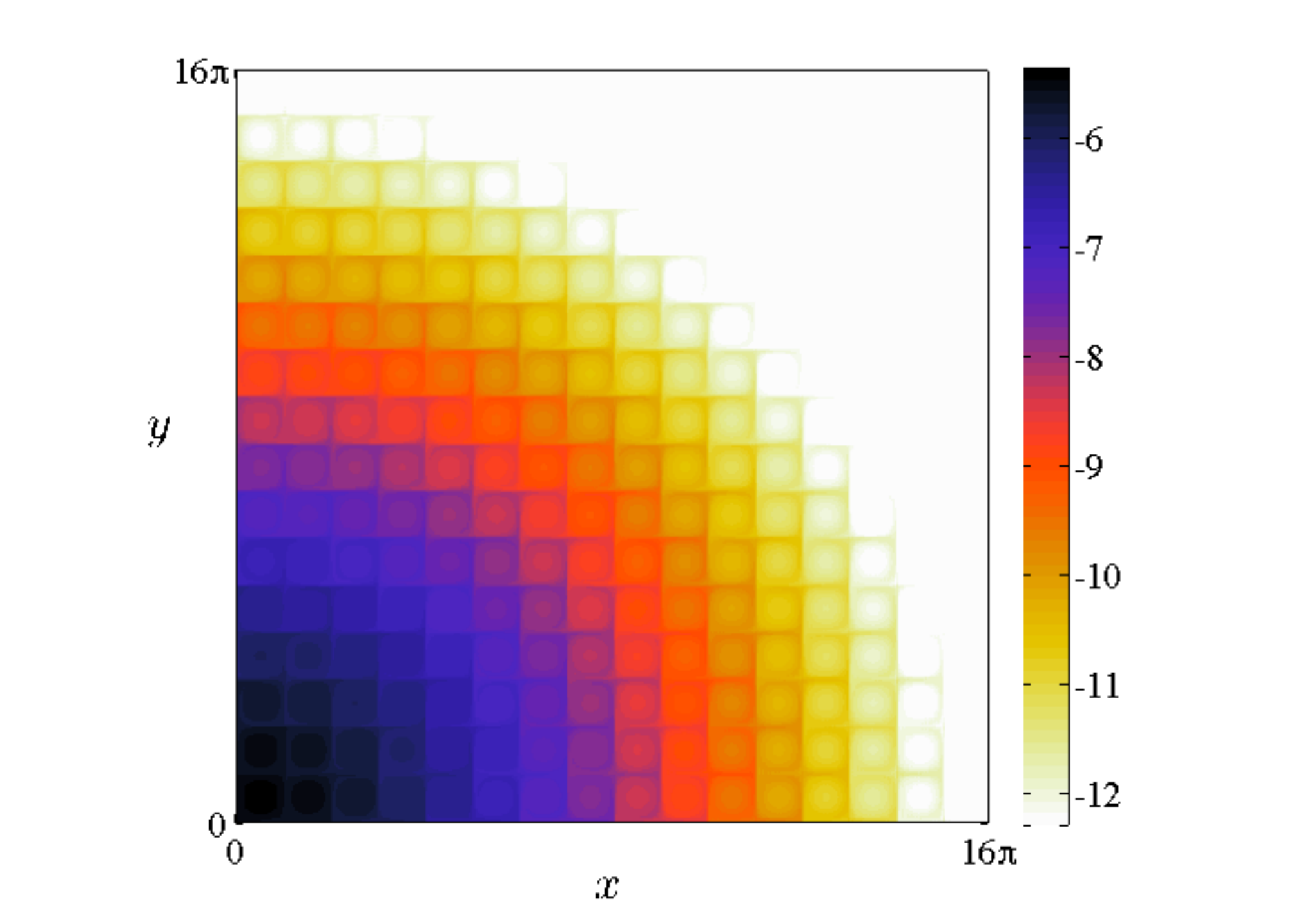}
\caption{(Colour online.) Concentration (in logarithmic scale)  at times $t=2$ (left) and $t=4$ (right) of a scalar released in the central cell of a cellular flow with $\Pe=250$.}
\label{fig:simu250}
\end{center}
\end{figure}

We focus our attention on the cellular flow
\beq \lab{cellflow1}
\bu(x,y)=(-\partial_y \psi,\partial_x \psi) \inter{with} \psi= - \sin x \sin y.
\eeq
This flow, with period $2\pi$ in both the $x$- and $y$-direction, consists of a regular array of cells in which the fluid is rotating alternatively clockwise and counterclockwise along closed streamlines; see Figure \ref{fig:cellpicturelarge}. It has received a great deal of attention, most of it devoted to the properties of the effective diffusivity that can be computed by homogenisation, especially in the limit of large P\'eclet number;  see \citet[][\S2]{majd-kram} for a review, and  \citet{novi-et-al} and \citet{gorb-et-al} for more recent references.

To illustrate the dispersion of a passive scalar in this flow, we show in Figures \ref{fig:simu1}--\ref{fig:simu250} the concentration field obtained by solving numerically the advection--diffusion equation \eqn{ad-dif} for $\Pe=1$ and $\Pe=250$. Only the first quadrant is shown since the field has a four-fold symmetry. For $\Pe=1$, molecular diffusion plays a major part across the domain, leading to a smooth evolution, with only some modulations in the form of diagonal bands in the central sector of the quadrant and of cells located near the coordinate axes. For $\Pe=250$, advection dominates, resulting in an apparent finite propagation speed and the obvious mark of the flow structure on the scalar field. The importance of the separatrices, around which boundary layers of high concentrations are established, is clear.
As the distance from the origin increases, there is gradual change in the scalar distribution within the cells, from almost uniform near the origin to essentially zero at large distance. This feature is discussed briefly below and fully elucidated in Part II.

 \begin{figure}
\begin{center}
\includegraphics[height=5cm]{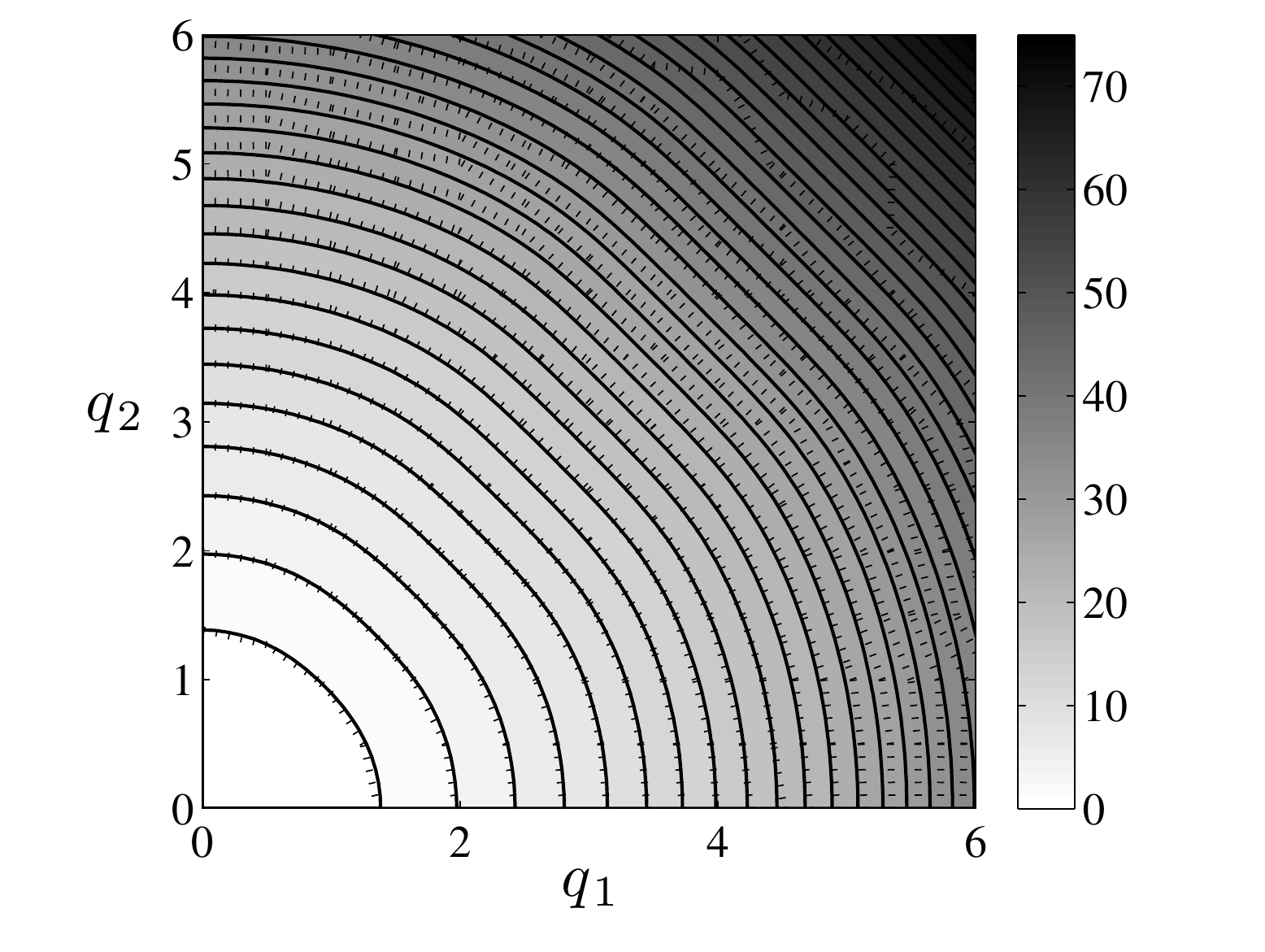} \hfill
\includegraphics[height=5cm]{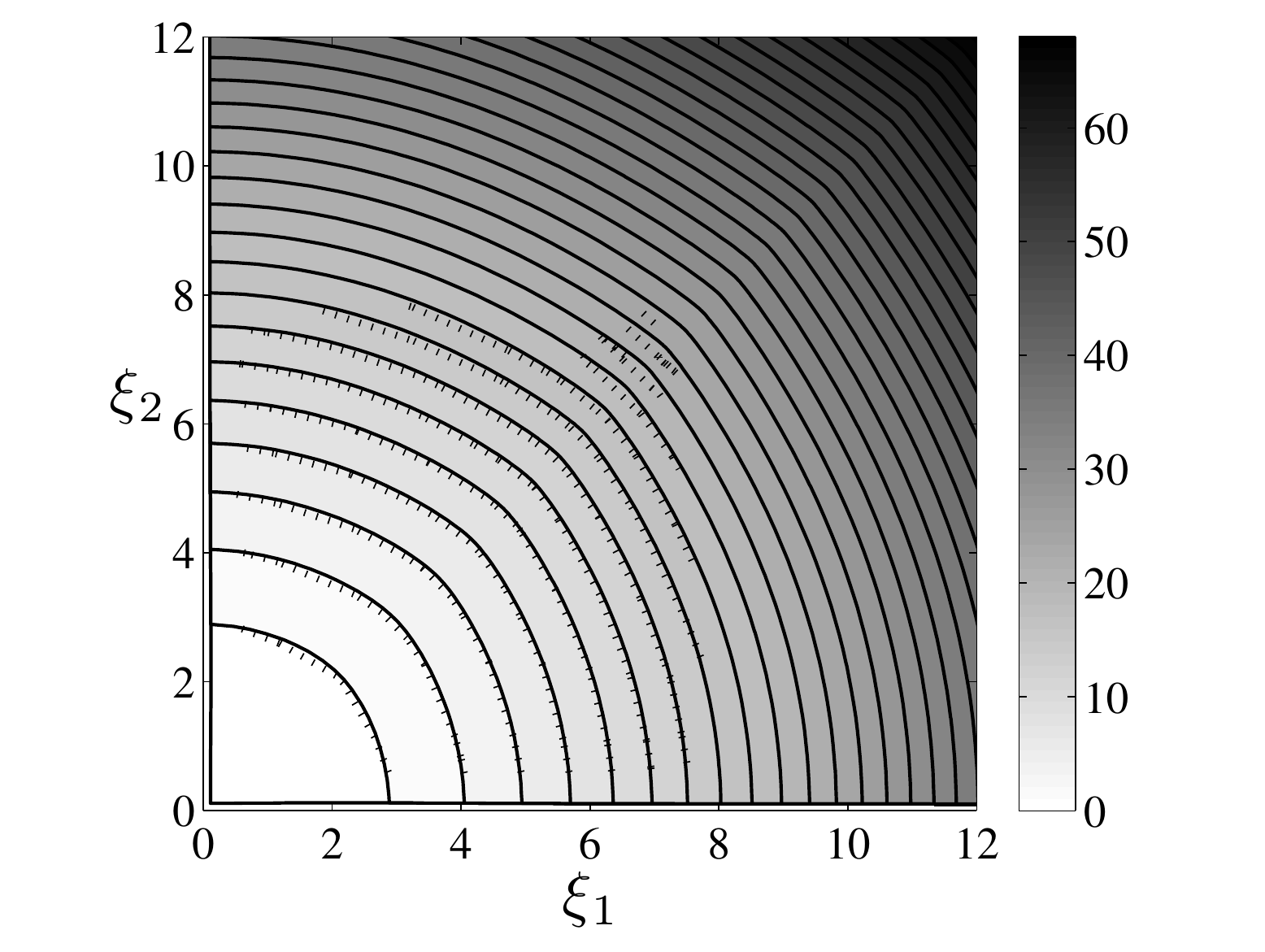} 
\caption{Left: $f$ as a function of $\bq$ for the cellular flow with $\Pe=1$. The solid contours and shading have been obtained by solving the eigenvalue problem \eqn{eig1} numerically, the dotted contours by Monte Carlo simulations with importance sampling ($10^5$ realisations for each value of $\bq$). Right: corresponding rate function $g$ as a function of $\bq$ obtained by Legendre transforming the results on the left. Note that the noise in the Monte Carlo results lead to an estimate of $g$ that is reliable in a restricted range of $\bxi$.}
\label{fig:cellf2dPe1}
\end{center}
\end{figure} 

Let us now turn to the predictions of large-deviation theory. 
We have developed a code for the numerical solution of the eigenvalue problem \eqn{eig1} for \eqn{cellflow1}. This relies on a straightforward finite-difference discretisation and on the matlab  routine `eigs' for the solution of the resulting matrix eigenvalue problem. The convergence of the algorithm requires a good first guess for the eigenvalue; since we are interested in obtaining $f(\bq)$ for a range of $\bq = (  q_1 , q_2 )$, the code performs an iteration over $q_1$ and $q_2$, using at each step the previous value of $f(\bq)$ as its first guess. Since $f$ satisfies the obvious symmetries $f(\pm q_1,\pm q_2)=f(q_1, q_2)$, we concentrate on the first quadrant of the $\bq$-plane. The symmetry $f(q_1,q_2)=f(q_2,q_1)$ can also be exploited. 

The left panel of Figure \ref{fig:cellf2dPe1} shows the numerical approximation to $f$ obtained using this code for $\Pe=1$. It is compared with the result of a Monte Carlo estimate which relies on the importance-sampling algorithm described in Appendix \ref{sec:resamp}. In addition to confirming the validity of the large-deviation approximation and of the numerical implementation, the figure illustrates general qualitative features of $f$. For small $|\bq|$, $f$ is approximately isotropic, consistent with the result of homogenisation theory which predicts a diagonal effective diffusivity tensor. For $|\bq|$ of order-one or larger, however, $f$ is anisotropic, taking smaller values along the axes $\bq=|\bq|(1,0)$ and  $\bq=|\bq|(0,1)$ than along the diagonal $\bq=|\bq|(1,1)/\sqrt{2}$. Physically, this implies that dispersion is slower along the axis than along the diagonal. This can be understood by considering the streamline geometry: continued advection along one of the axes requires particles to also meander in the perpendicular direction, resulting in a decrease in average speed by a factor $1/2$; by contrast, advection along the diagonal happens in staircase-like fashion which decreases the speed by a factor $1/\sqrt{2}$.
That motion along the diagonal is faster is also apparent in the rate function $g(\bxi)$ obtained by Legendre transform and shown on the right panel of Figure \ref{fig:cellf2dPe1}: when $|\bxi|$ is not small, the contours of $g$, which directly correspond to concentration contours, are anisotropic with the larger scalar concentrations along the diagonal. 

 \begin{figure}
\begin{center}
\includegraphics[height=5.7cm]{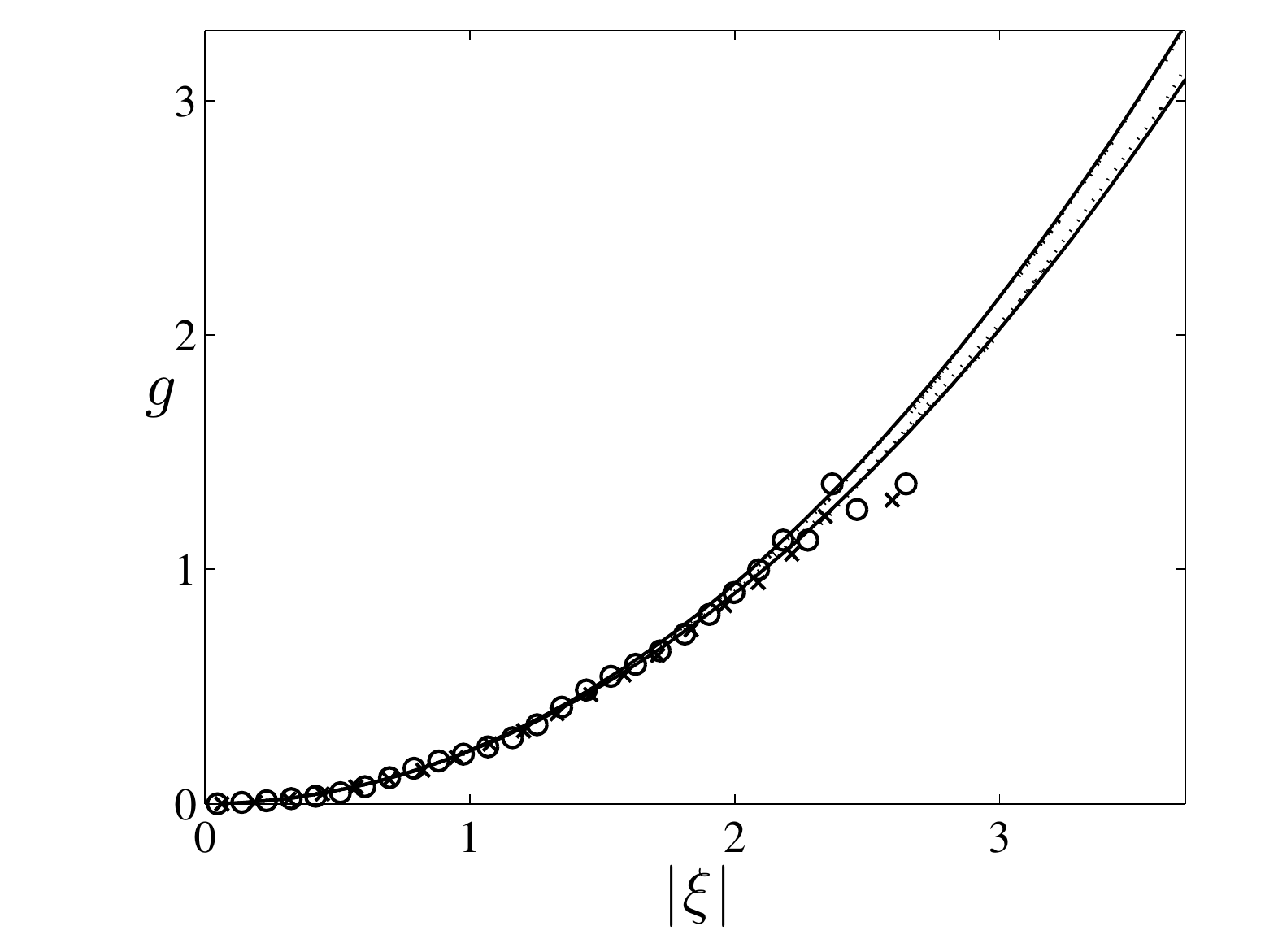} 
\caption{Rate function $g$ as as function of $|\bxi|$ for the cellular flow with $\Pe=1$. The curves have been obtained by Legendre transforms of $f$ computed by solving the eigenvalue problem (solid curves) and Monte Carlo simulation ($10^4$ realisations for each $\bq$, dotted curves); the symbols have been obtained from a direct Monte Carlo estimation of the particle position pdf ($4 \times 10^7$ realisations). The two pairs of curves and two types of symbols correspond to $\bxi=|\bxi|(1,0)$ (steeper curves and circles) and $\bxi=|\bxi| (1,1)/\sqrt{2}$ (shallower curves and squares).}
\label{fig:cell3ways}
\end{center}
\end{figure}

A direct Monte Carlo estimate of $g$ --- as opposed to the indirect estimate deduced from Legendre transforming the Monte Carlo approximation to $f$ --- proves difficult to compute reliably. Figure \ref{fig:cell3ways} illustrates this: even for a large number of realisations of 
$4 \times 10^7$, the direct Monte Carlo approach only provides a valid approximation for $|\bxi| \lesssim 2.5$, in range where $g$ remains roughly isotropic. Attempts at implementing importance sampling in a manner analogous to that used for shear flows and described in Appendix \ref{sec:girs} did not lead to significant improvements in the estimation of $g$ in this direct manner. A conclusion, therefore, is that a more efficient Monte Carlo approximation to $g$ is achieved by sampling $f$ and taking a Legendre transform. Of course, for this problem the most efficient method for obtaining $f$ and $g$ remains the numerical solution of the eigenvalue problem \eqn{eig1}.

\begin{figure}
\begin{center}
\begin{tabular}{ccc}
\hspace{-1cm} \includegraphics[height=4.8cm]{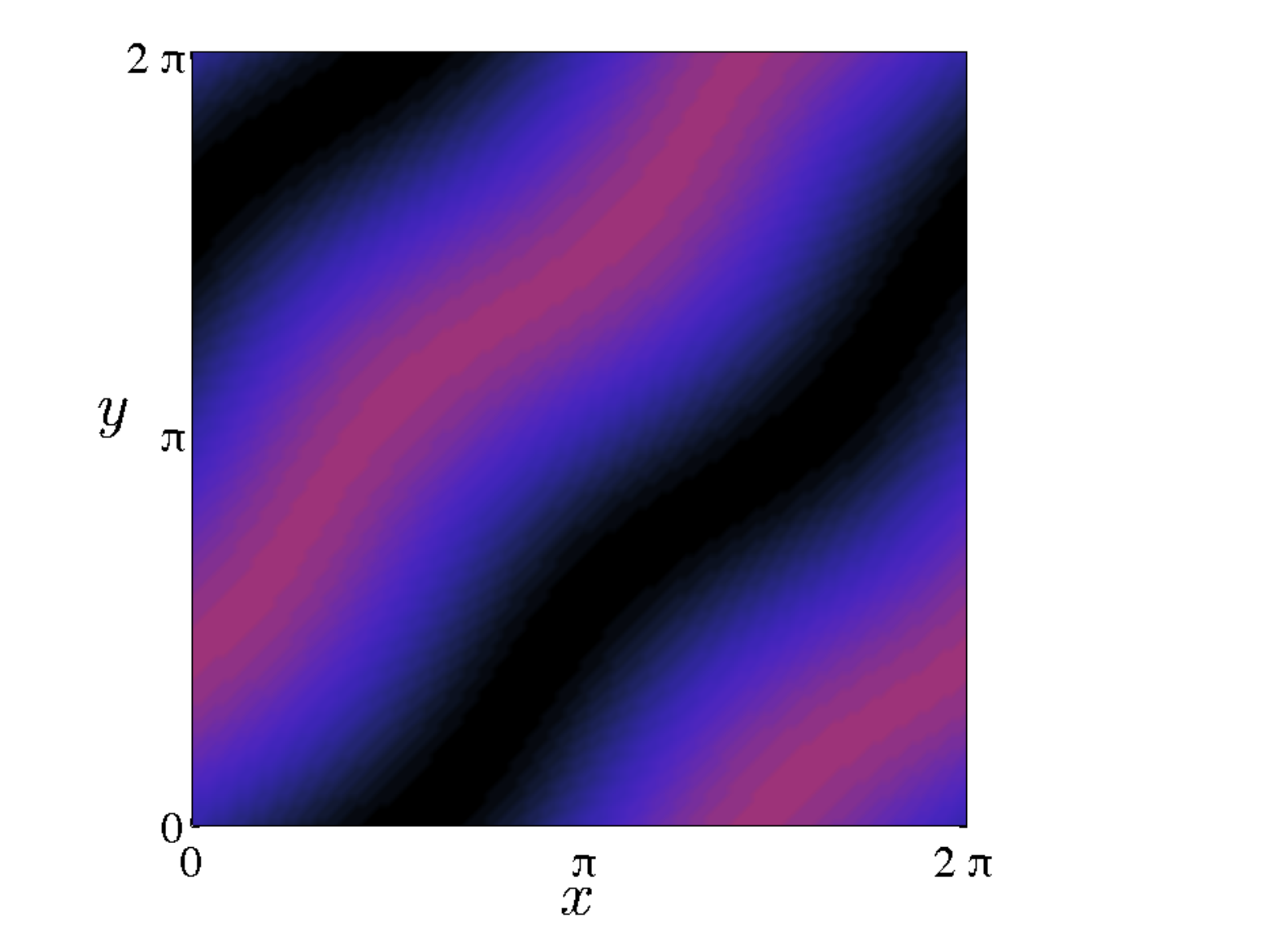} &  \hspace{-1.8cm}
\includegraphics[height=4.8cm]{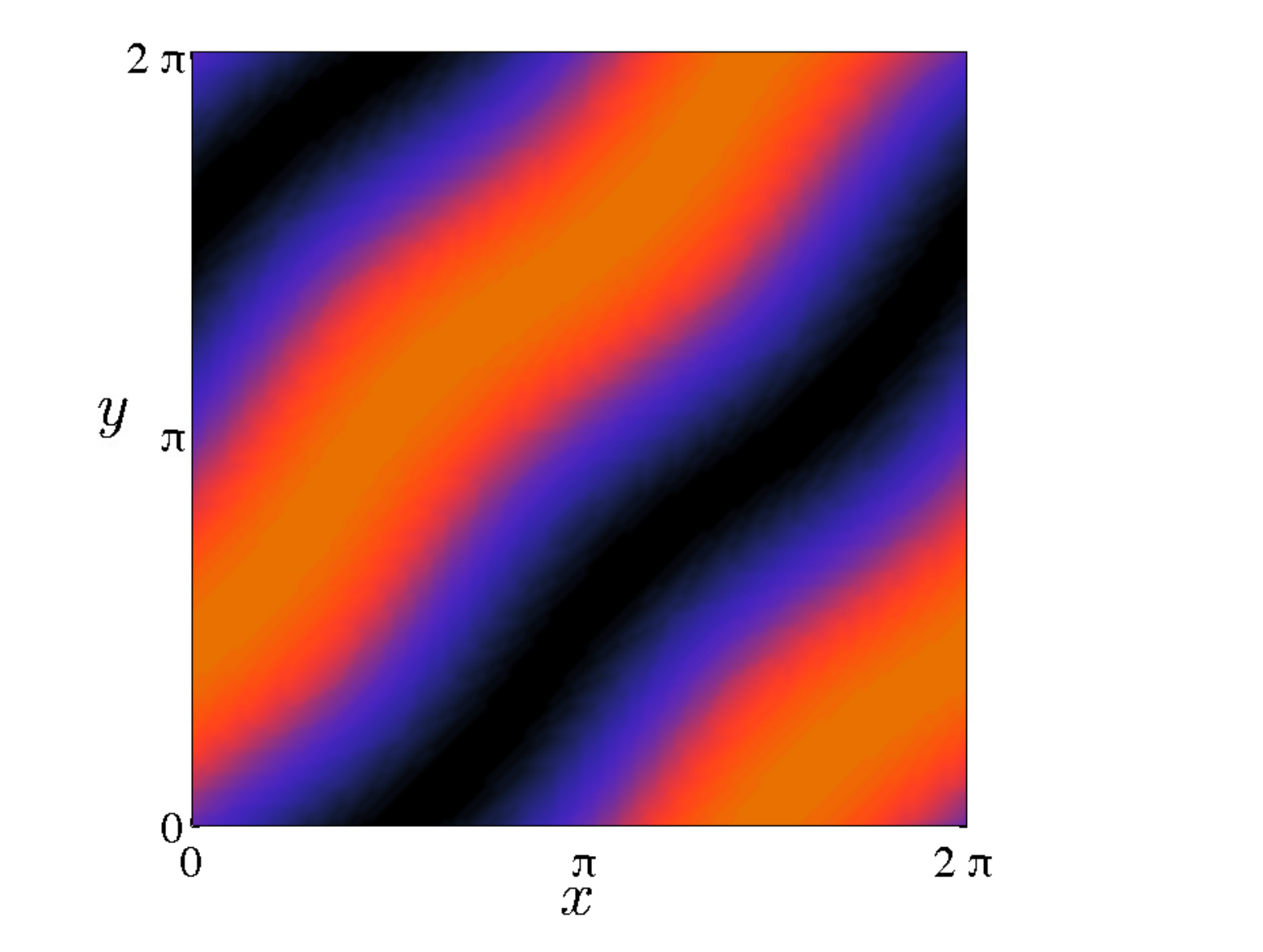} & \hspace{-1.6cm}
\includegraphics[height=4.8cm]{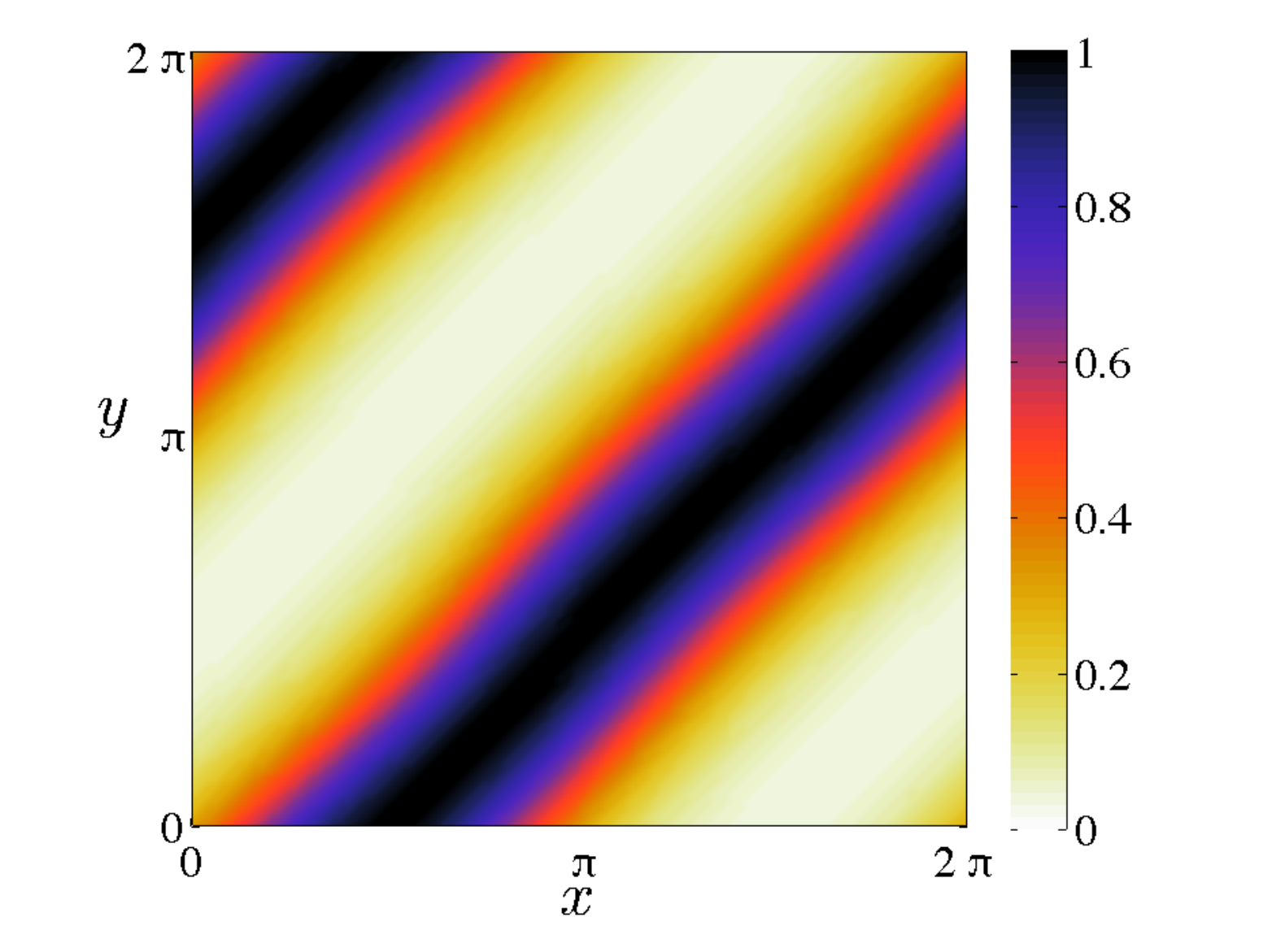}
\end{tabular}
\caption{(Colour online.) Eigenfunctions for $\Pe=1$ and $q_1=q_2=0.5$ (left), $1$ (middle) and $5$ (right). The eigenfunctions have been normalised to have maximum value $1$ and plotted using the same colour scale shown on the right.} \label{fig:efunctionPe1}
\end{center}
\end{figure}

It is interesting to examine the eigenfunctions $\phi$ associated with the eigenvalue $f(\bq)$ for given $\bq$ since these provide the structure of the scalar field at position $\bxi t = \nabla_{\bq} f(\bq) t$ (with $f$ convex so that $\bq$ can be interpreted as a proxy for $\bxi$). Figure \ref{fig:efunctionPe1} shows the eigenfunctions obtained by numerical solution of the eigenvalue problem for three values of $q_1=q_2=|\bq|/\sqrt{2}$. 
For small $|\bq|$ and hence small $|\bxi|$, $\phi$ is essentially constant over the whole cell, with only small modulations. This is consistent with the perturbative treatment of the eigenvalue problems for $|\bq| \ll 1$ and  $|\bxi| \ll 1$, amounting to homogenisation, which indicates that $\phi = 1 + O(|\bq|)$. As $|\bq|$ and $|\bxi|$ increase, the modulations, in the form of diagonal stripes, increase in amplitude so that, for large $|\bxi|$, $\phi$ is close to zero  in wide stripes. The form of the eigenfunctions depends on the angle of $\bq$, of course, and for $q_1=0$ or $q_2=0$ for instance, corresponding to dispersion along the $x$ and $y$ axes, they have a  have a cellular rather than banded structure (not shown). The structure of the eigenfunctions is consistent with the concentration field shown in Figure \ref{fig:simu1}. To see this, recall that the concentration depends on both $\phi$ and on the rate function $g$; across a single cell, the latter varies slowly and can be approximated by a Taylor expansion, leading to the spatial dependence $\phi(\bx,\bq) \exp(\bq \cdot \bx)$, since $\nabla g = \bq$. For large $|\bq|$, the dominant effect is the exponential decay of the concentration in the direction of $\bq$, with the form of $\phi$ introducing the banded modulations observed in Figure \ref{fig:simu1}.

%

 \begin{figure}
\begin{center}
\includegraphics[height=5.7cm]{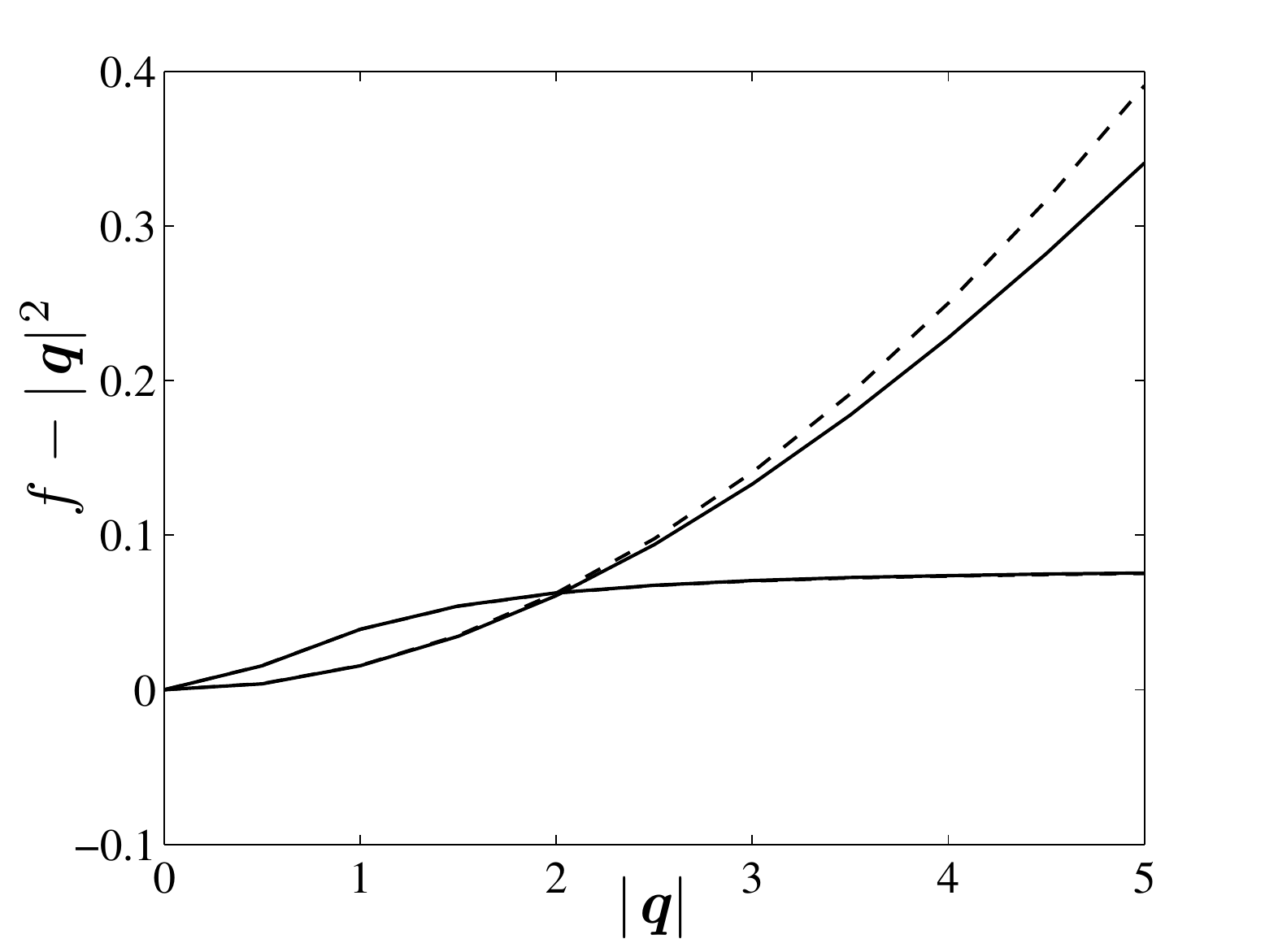} 
\caption{Correction $f-|\bq|^2$ as a function of $|\bq|$ for the cellular flow with $\Pe=1/4$ and for $\bq=|\bq| (1,1)/\sqrt{2}$ (rapidly growing curves) and $\bq=|\bq| (1,0)$ (other curves, values multiplied by 10). The exact result (solid) is compared with the small-$\Pe$ approximation (dashed).}
\label{fig:smallPf}
\end{center}
\end{figure} 

Some insight into the large-deviation behaviour of cellular flows can be gained by considering the regime $\Pe \ll 1$ corresponding to weak advection. The effective diffusivity in this limit was computed by \citet[][\S7]{moff83} and \citet{sagu-hors} who obtained (in our notation) the approximation $\keff = 1 +\Pe^2/8 + O(\Pe^4)$. The generalisation to the large-deviation regime is straightforward and described in Appendix \ref{sec:smallPe}. It leads to the asymptotic approximation
\beq \lab{smallPe}
f(\bq)=q_1^2 + q_2^2 + \frac{\Pe^2}{8} \frac{q_1^2 + q_2^2 + q_1^4 + 6 q_1^2 q_2^2 + q_2^4}{1+2(q_1^2+q_2^2) + (q_1^2-q_2^2)^2} + O(\Pe^3)
\eeq
whose small-$\bq$ limit is consistent with the effective diffusivity just quoted. This approximation is tested against numerical results in Figure \ref{fig:smallPf} which shows the correction $f(\bq)-|\bq|^2$ to purely diffusive behaviour for $\Pe=1/4$. The figure confirms the validity of \eqn{smallPe}; it  also shows that dispersion is fastest along the diagonal, as noted for $\Pe=1$. The $O(\Pe^2)$ correction to $f$ behaves in fact very differently for $q_1=q_2$ than it does for $q_1\not=q_2$: whereas is is bounded as $\bq \to \infty$ for $q_1 \not=q_2$, it grows quadratically for $q_1=q_2$ in a manner that suggests that \eqn{smallPe} is not uniformly valid. Eq.\ \eqn{smallPe} shows immediately that the difference in behaviour stems from the fact that the denominator of the $O(\Pe^2)$ term is  quadratic for $q_1=q_2$ but quartic, like the numerator, otherwise. This is the manifestation of a phenomenon that can be captured by a large-$|\bq|$ asymptotic analysis which we do not present here. Briefly, this analysis reveals the direction $q_1=q_2$ to be singular for the flow \eqn{cellflow1} in that the correction to the diffusive behaviour $f(\bq) \sim |\bq|^2$ is $O(|\bq|)$ in this direction while it is $O(1)$ in all other directions. Flows with more complicated spatial structures than \eqn{cellflow1} have other singular directions so that we expect the dependence of $f(\bq)$ on the direction of $\bq$ to be very complicated.

We conclude our discussion of cellular flows by briefly considering the large-$\Pe$ regime. This is the regime that has received most attention in the now extensive literature on effective diffusivity for cellular flows. Starting with \citet{chil79}, several authors have applied a boundary-layer analysis to the cell problem of homogenisation to conclude that $\keff \propto \Pe^{1/2}$ in this case \citep[see][]{shra87,rose-et-al}, with \citet{sowa87} deriving an explicit expression for the proportionality constant. Part II of the present paper is devoted to a detailed asymptotic treatment of the large-deviation eigenvalue problem for $\Pe \gg 1$ which recovers and extends this result. Here we only discuss some qualitative properties of the solution derived numerically.

\begin{figure}
\begin{center}
\includegraphics[height=5cm]{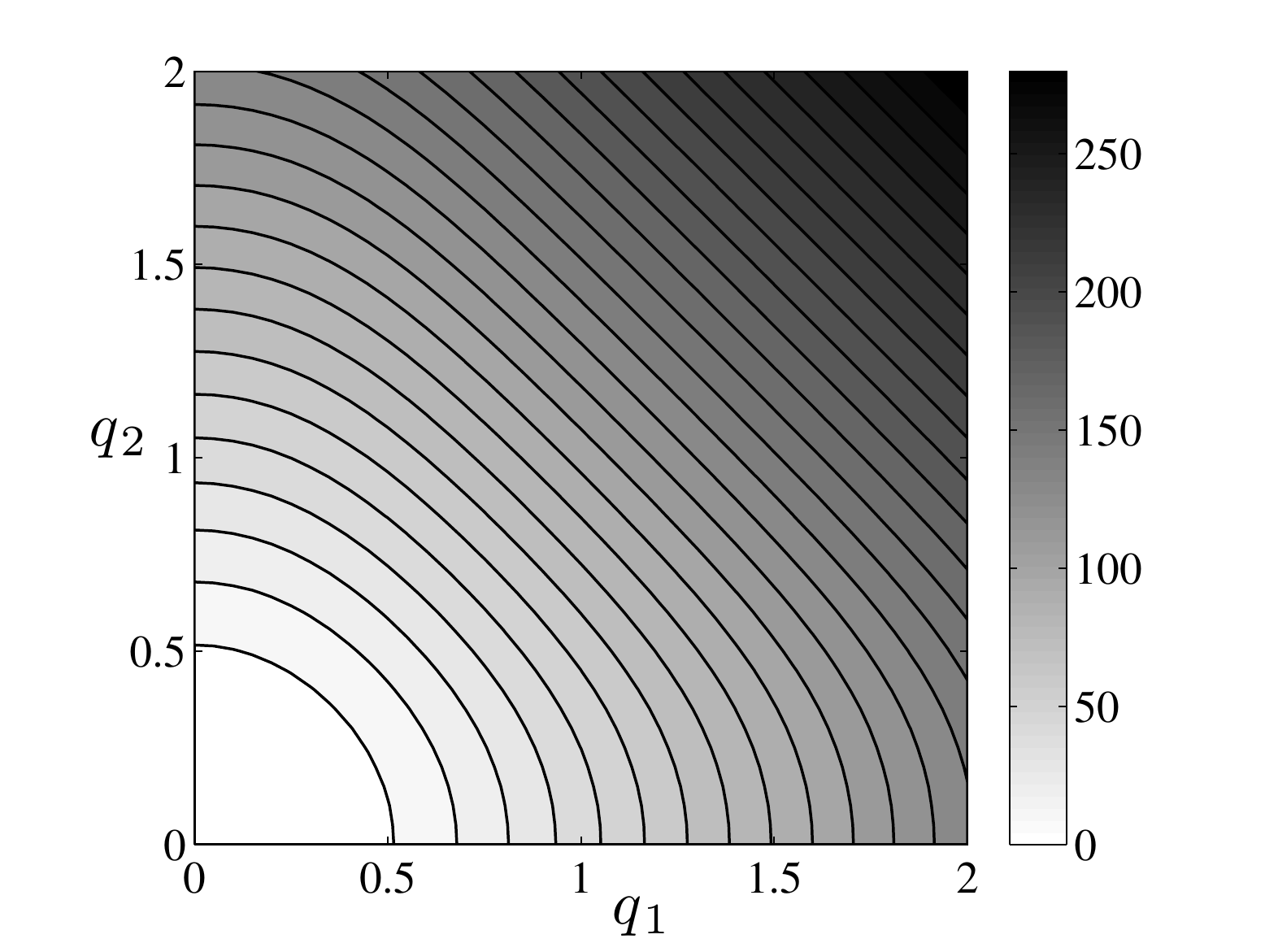} \hfill
\includegraphics[height=5cm]{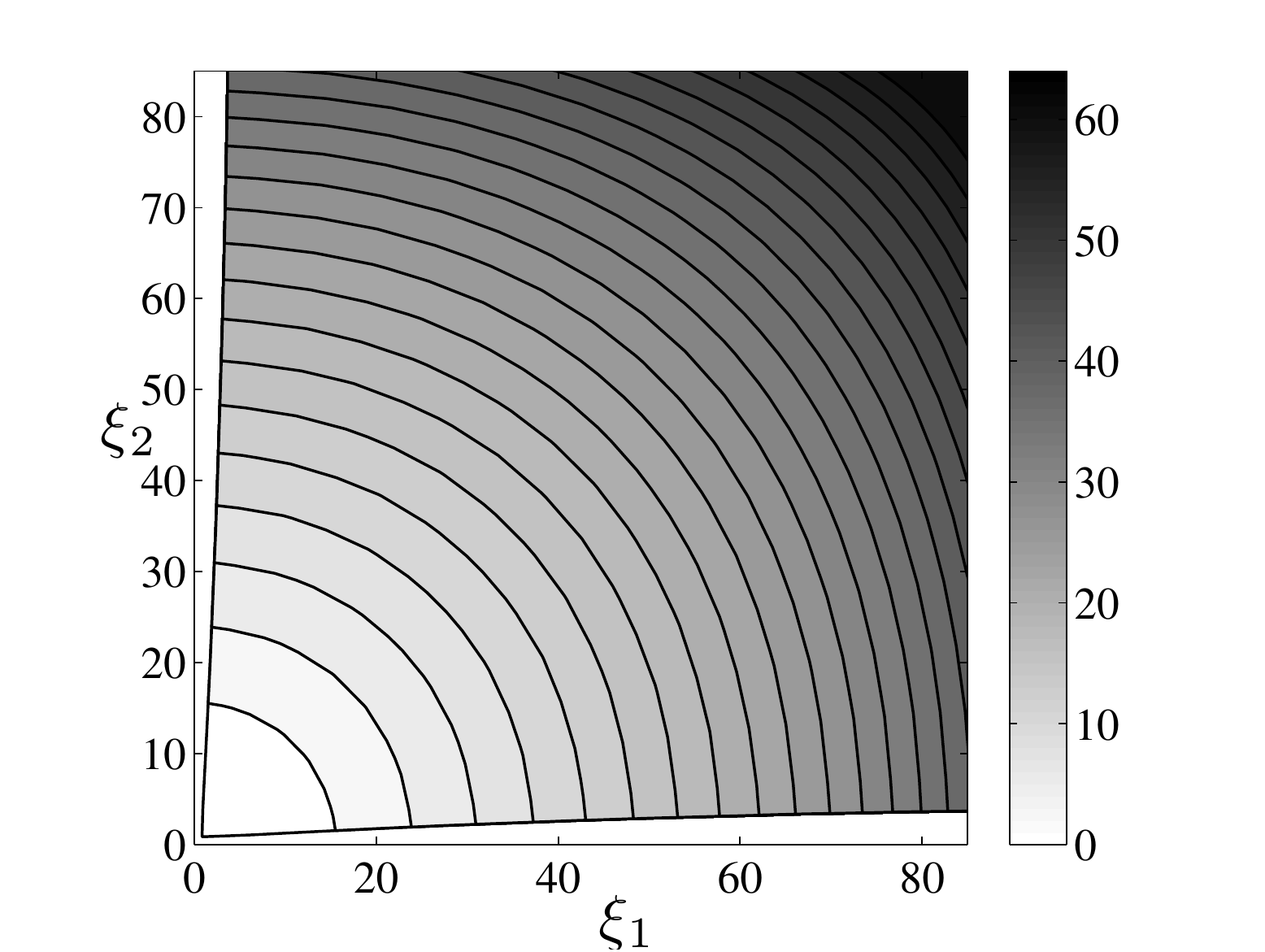} 
\caption{Left: $f$ as a function of $\bq$ obtained by solving the eigenvalue problem \eqn{eig1} for the cellular flow with $\Pe=250$.  Right: rate function $g$ deduced by Legendre transform.}
\label{fig:cellf2dPe250}
\end{center}
\end{figure}

Figure \ref{fig:cellf2dPe250} shows $f$ and $g$ obtained by numerical solution of the eigenvalue problem and Legendre transform for $\Pe=250$. The anisotropy for $|\bq| \gtrsim 1$ observed for $\Pe=1$ is stronger for this large-$\Pe$ case: there is a clear suggestion that the contours of $f(\bq)$ tend to straight lines (corresponding to $f$ being a function of $|q_1|+|q_2|$) for $|\bq| \gg 1$; correspondingly, $g(\bxi)$ depends on  $\max(|\xi_1|,|\xi_2|)$ for $|\bxi|\gg 1$.

\begin{figure}
\begin{center}
\begin{tabular}{ccc}
\hspace{-1cm} \includegraphics[height=4.8cm]{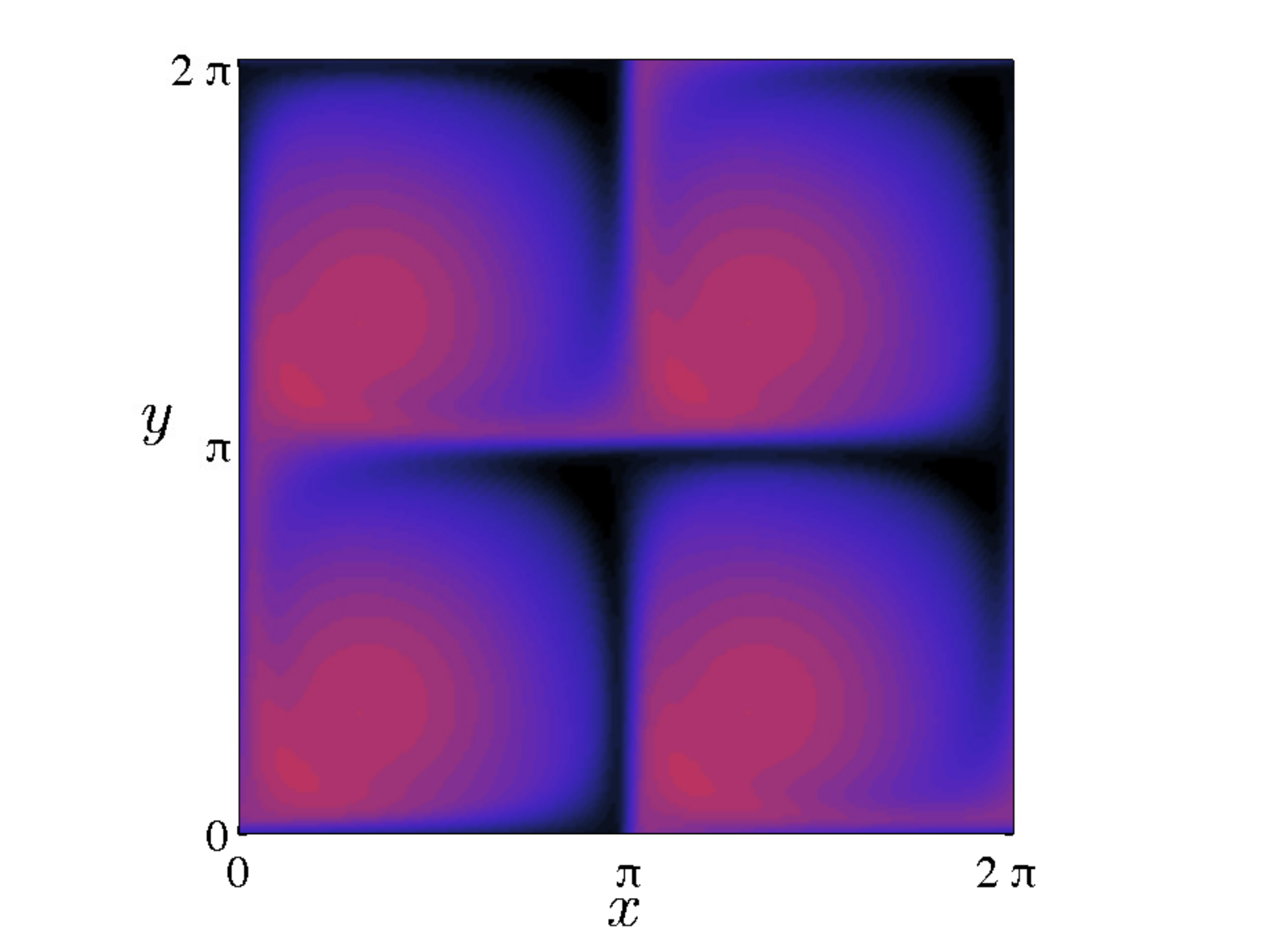} &  \hspace{-1.8cm}
\includegraphics[height=4.8cm]{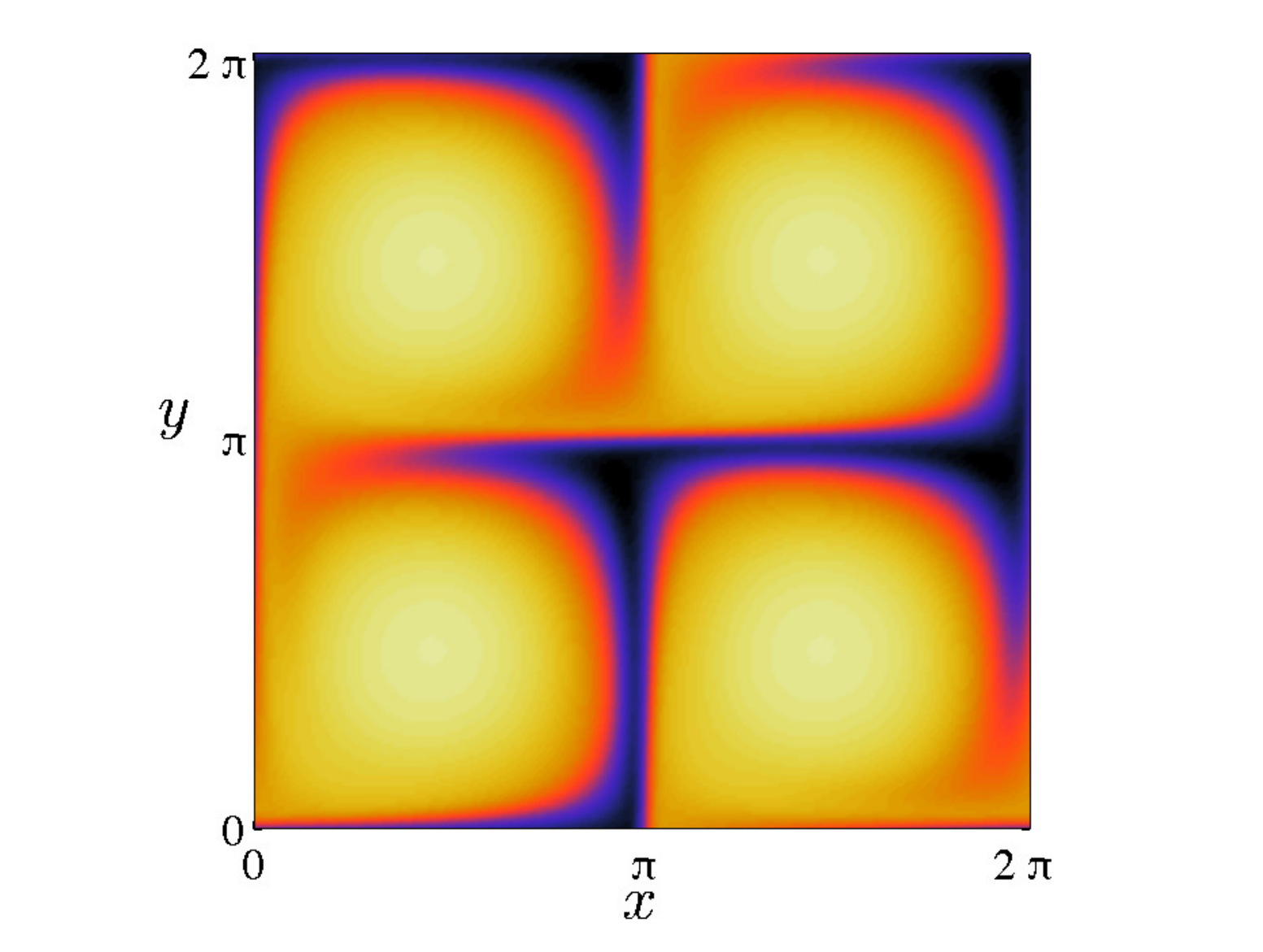} & \hspace{-1.6cm}
\includegraphics[height=4.8cm]{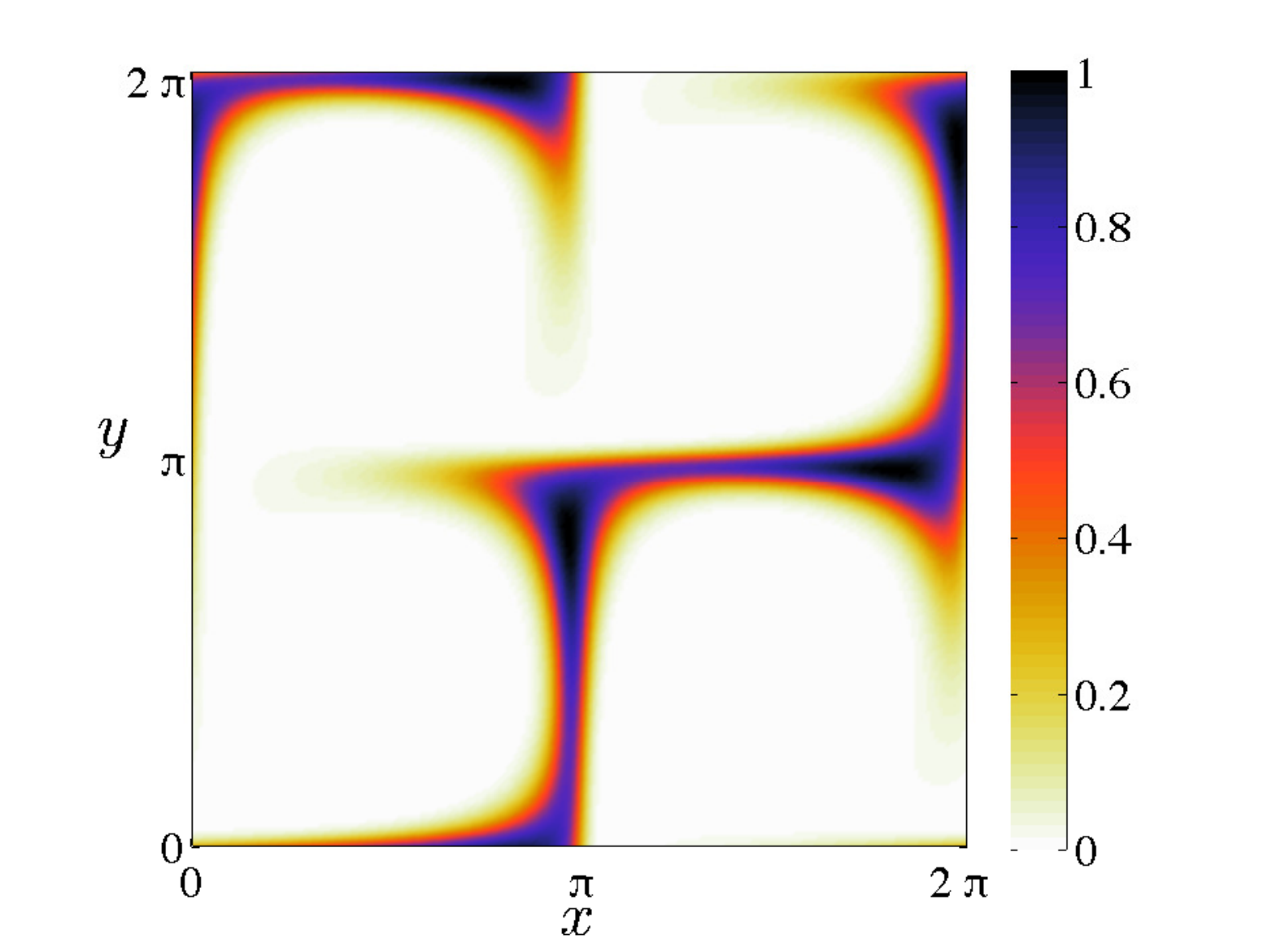}
\end{tabular}
\caption{(Colour online.) Eigenfunctions for $\Pe=250$ and $q_1=q_2=0.1$ (left), $0.25$ (middle) and $1$ (right), corresponding to $\xi_1=\xi_2=4.2,\, 20.5$ and $88.1$. The eigenfunctions have been normalised to have maximum value $1$ and plotted using the same colour scale shown on the right.} \label{fig:efunctionPe250}
\end{center}
\end{figure}
 
The eigenfunctions of \eqn{eig1} shown in Figure \ref{fig:efunctionPe250} for three different values of $q_1=q_2$ illustrate different regimes of  dispersion that arise at increasingly larger distances from the scalar-release point. For small $|\bq|$ and hence for small $|\bxi|$, $\phi$ is almost uniform: a gentle $O(|\bq|)$ gradient in the cell interiors is compensated by a rapid change in boundary layers that appear around the separatrices in agreement with the homogenisation treatment. For larger $\bq$ and $|\bxi|$, $\phi$ inside the cell is no longer close to uniform; instead, it is approximately constant along streamlines but varies across streamlines, from small values at the centre to large values near the separatrices.  Again, boundary layers around the separatrices ensure periodicity. Finally, for large $|\bq|$ and $|\bxi|$, $\phi$ is close to zero in the cell interiors and the scalar is confined within boundary layers.  This qualitative description of the eigenfunctions is consistent with the evolution of the scalar field shown in Figure \ref{fig:simu250}; it is supported by the asymptotics results reported in Part II.

\section{Discussion} \label{sec:disc}

This paper discusses the statistics of passive scalars or particles dispersing in fluids under the combined action of advection and molecular diffusion. It shows how large-deviation theory provides an approximation to the scalar concentration or particle-position pdf in the large-time limit. This approximation, expressed in terms of the rate function $g(\bxi)$, is valid in the tail of the distribution as well as in the core; it considerably generalises the more usual diffusive approximation which characterises the dispersion by a single effective-diffusivity tensor. The rate function is deduced from the solution of the generalised cell problem \eqn{eig1}, a one or two-parameter family of eigenvalue problems that generalise the cell problem solved when computing the effective diffusivity using the method of homogenisation. 

The application to shear flows reveals features of the dispersion that are not captured by the standard theory of shear dispersion initiated by \citet{tayl53}. In particular, it shows that the diffusive approximation dramatically overestimates scalar concentrations far away from the centre of mass. The reason for this is that the mechanism underlying shear dispersion---the interaction between shear and cross-stream molecular diffusion---leads to along-flow dispersion with a finite speed, namely the maximum flow speed. The non-zero concentrations beyond the limits imposed by this finite speed are entirely attributable to molecular diffusion and  thus controlled by  molecular rather than effective diffusivity.\footnote{Molecular diffusion itself, with its infinite propagation speed, is of course only a model for Brownian motion; more sophisticated models with finite propagation speeds such as the telegraph equation can be developed (e.g., \citealt{zaud06}; see \citealt{kell04} for connections with large deviations).} At intermediate distances from the centre of mass, however, the non-diffusive effects can  in some cases increase and in some cases decrease dispersion. This can be detected in some of the results for standard shear flows displayed in Figure \ref{fig:shearall} or be deduced from the order-by-order corrections to the diffusive approximation discussed in \S\ref{sec:hom}.

Our analysis of spatially periodic flows and, in particular, of the classical cellular flow further demonstrates the benefits of large-deviation theory over homogenisation and the resulting diffusive approximation. The anisotropy of the dispersion in this flow, for instance, although a clear consequence of the streamline arrangement, is overlooked by the diffusive approximation but quantified by large deviation. As for shear flows, there is also a finite speed effect for the dispersion in cellular flow; this is more subtle and is elucidated in Part II which devoted to a detailed analysis to the large-$\Pe$ limit.

The differences between the diffusive and large-deviation approximations for the scalar concentration are  significant at large enough distances away from the centre of mass of the scalar. Since the concentration at such distances is  small, large deviation applied to problems involving purely passive scalars is of practical importance in situations where low concentrations matter, as would be the case, for instance, for very toxic chemicals. In such applications the logarithm of the concentration is often a relevant measure of the chemical's impact; it is read off from the rate function since $\log C \sim - t  g(\bxi)$. 
As mentioned in \S1, for scalars that are reacting, the properties of dispersion at large distances embodied in $g$  can be critical in determining the main features of the scalar distribution. This was made explicit in the work of \citet{gart-frei} and \citet{frei85} which relates the speed of propagation of fronts for scalars experiencing F-KPP-type reactions to the rate function $g(\bxi)$ characterising passive dispersion. Following from this relationship, the results of the present paper and of Part II can be used to predict front speeds in a range of shear and periodic flows. We will report elsewhere the novel predictions that can be obtained in this manner \citep{tzel-v14b,tzel-v14a}.

We conclude by remarking that the large-deviation treatment of scalar dispersion can be extended to a class of flows much broader than that considered in the present paper. Dispersion in time-periodic flows,  random flows and turbulent flows can also be characterised by a rate function to improve on the approximation provided by  effective diffusivity.  In the time-periodic case an extension of the theory discussed in \S2 is straightforward: the eigenfunction $\phi$ in \eqn{largedevi} should depend on $t$ as well as on $\bx$ and $\bxi$, leading to an additional term $\partial_t \phi$ in the eigenvalue problem \eqn{eig1} and to the  requirement that $\phi$ be time periodic which determines the eigenvalue $f$. In the random case, under the assumption of homogeneous and stationary statistics for $\bu(\bx,t)$, $f$ is determined by the analogous requirement that $\phi$,  a random function, be homogeneous and stationary. Implementing this requirement is not straightforward, however, and  Monte Carlo methods with importance sampling of the types described in Appendix B may be best suited for the computation of the rate function.

\smallskip

\noindent
\textbf{Acknowledgments.} JV acknowledges support from grant EP/I028072/1 from the UK Engineering and Physical Sciences Research Council.

\appendix

\section{Small-$|\bq|$ expansion for shear flows} \label{sec:exp}

It follows from the scaled large-deviation form of $C$ for shear flows \eqn{Cdisp} that
\[
\partial_t C \sim (g' \xi - g) C = f(q) C  \inter{and} \partial^n_x C \sim (-\Pe^{-1} g')^n C = (-\Pe^{-1} q)^n C.
\]
In these expressions, $q$ is related to $\xi=\Pe^{-1}x/t$ by $\xi=f(q)$ and factors
$1 + O(t^{-1})$ describing the error in the WKB-like expansion \eqn{Cdisp} are omitted.
Thus if we write
\beq \lab{exp}
f(q) \sim \sum_{n=1}^N \alpha_n q^n,
\eeq
an equation for $C$ follows in the form
\[
\partial_t C \sim \sum_{n=1}^N (-\Pe)^n \alpha_n \partial_x^n C.
\]
The solution to this equation gives for $C$ a form similar to \eqn{Cdisp} with $g$ approximated by the Legendre transform of the $N$-term Taylor expansion of $f(q)$ at $q=0$. In particular, truncating at $N=2$ gives the dispersive approximation with effective diffusivity \eqn{effdiff}.

The perturbative solution of \eqn{eig-disp} is straightforward: introducing \eqn{exp} and 
\[
\phi(y) = 1 + \sum_{n=1}^N q^n  \phi_n(y)
\]
into \eqn{eig-disp} and omitting the term in $\Pe^{-2}$ gives at the first three orders,
\beq \lab{3eqns}
\dt{^2 \phi_1}{y^2} = \alpha_1 - u, \quad \dt{^2 \phi_2}{y^2} = \alpha_2 + \alpha_1 \phi_1 - u \phi_1 \inter{and}  \dt{^2 \phi_3}{y^2} = \alpha_3 + \alpha_2 \phi_1 + \alpha_1 \phi_2 - u \phi_2.
\eeq
Integrating the first equation and using \eqn{zeroav} gives $\alpha_1=0$ and
\beq 
\dt{\phi_1}{y} = - \int_{-1}^y u(y') \, \d y'.
\eeq
An explicit expression for $\phi_1$ follows, which can be chosen such that $\langle \phi_1 \rangle = 0$.
Integrating the second equation in \eqn{3eqns} and using the above gives
\beq \lab{alpha2}
\alpha_2 = \langle u \phi_1 \rangle = \langle \left(\int_{-1}^y u(y') \, \d y' \right)^2 \rangle .
\eeq
Up to the factor $\Pe^2$, this is the effective diffusivity of Taylor and homogenisation theory. The function $\phi_2(y)$ can then computed explicitly and the condition $\langle \phi_2 \rangle = 0$ imposed.
Finally, integrating the third equation in \eqn{3eqns} gives
\beq \lab{alpha3}
\alpha_3 = \langle u \phi_2 \rangle = \langle u \phi_1^2 \rangle,
\eeq
in agreement with \citet{youn-jone}. Note that the analogue of \eqn{alpha2} for pipe flows is
\beq \lab{alpha2ax}
\alpha_2 = 2 \int_0^1 \left( \int_0^r r' u(r') \, \d r' \right)^2 \frac{\d r}{r}.
\eeq

\section{Monte Carlo computations} \label{sec:motecarlo}

\subsection{Resampled Monte Carlo} \label{sec:resamp}

We test the theoretical results by estimating the cumulant generating function from Monte Carlo simulations. This relies on solving \eqn{sde} for an ensemble of trajectories $\bX^{(k)}, \, k=1,\cdots,K$, then computing
\beq \lab{samp}
W_K(t) = \frac{1}{K} \sum_{k=1}^K w^{(k)}(t), \quad \textrm{where} \ \ w^{(k)}(t) = \e^{\bq \cdot \bX^{(k)}(t)},
\eeq
for fixed $\bq$. Since $W_K(t) \to \E \exp(\bq \cdot \bX)$ as $K \to \infty$, $f(\bq) \approx t^{-1} \log W_K(t)$ for $t$ and $K$ large.
 
When $\bq$ is small, this method provides a good estimate of $f(\bq)$ with $t$ moderately large, say $t=5$ or $10$. For $\bq$ of order one or large, obtaining even a crude estimate of $f(\bq)$ requires an exceedingly large number of realisations $K$.
This is because the cumulant generating function is determined by exponentially rare, hence difficult to sample, realisations whose weights $w^{(k)}(t)$ are exponentially larger than those of typical realisations.  To estimate $f(\bq)$ accurately with a reasonable number of realisations, it is necessary to use an importance-sampling method which concentrates the computational efforts on realisations that dominate \eqn{samp}. We have adopted a simple method based on Grassberger's \citeyearpar{gras97} pruning-and-cloning technique \citep[see also][]{gras02,tail-kurc07,v10} which we now describe.

Every few time steps in the numerical integration of \eqn{sde}, the current weight $w^{(k)}(t)$ of each realisation is compared to the average $W_K(t)$. If $w^{(k)}(t) > P W_K(t)$, where $P>1$ is a parameter of the method (typically chosen as $P=2$ or $3$), the realisation is cloned: an additional realisation $\bX^{(l)}$ is created and integrated forward from the initial condition $\bX^{(l)}(t)=\bX^{(k)}(t)$. The two clones  subsequently follow different trajectories, $\bX^{(l)}(t')\not=\bX^{(k)}(t')$ for $t'>t$ because they experience different Brownian motions. The statistics of $W_K(t)$ are left unchanged provided that the weight of the cloned realisations is divided by $2$, that is,  the weights $w^{(k)}(t)$ in \eqn{samp} are multiplied by additional factors of $1/2$ for each cloning experienced by realisation $k$. If $w^{(k)}(t) < W_K(t)/P$, on the other hand, the realisation is pruned: it is killed with probability $1/2$ and, if surviving, its weight $w^{(k)}(t)$ is multiplied by $2$. To keep the number of realisations $K$ constant, random realisations are either cloned or killed. We have implemented a slight extension of the method described in which the number of clones for realisations with $w^{(k)}(t) > P W_K(t)$, is $\lfloor w^{(k)}(t) / W_K(t) \rfloor + 1$.  

The resampling steps make the method very efficient, and the results reported in the paper typically required a few minutes of computation on a modest desktop computer. Crucial to this efficiency is the fact that the cloning-pruning process tailors the ensemble of realisations to a particular value of $\bq$ by selecting those which dominate $\E\exp(\bq \cdot \bX)$.

\subsection{Modified dynamics} \label{sec:girs}

The rate function $g$ can be estimated directly by Monte Carlo simulations, using a binning procedure to approximate $C$. This is of course highly inefficient for the parts of $g$ away from its minimum $\bxi_*$ since these are controlled by exponentially rare realisations which are poorly sampled. One way of remedying this is to integrate a modified dynamics following the importance-sampling technique discussed in \citet{mils95}. For shear flows, we have adopted the following approach.
The modified dynamics, denoted by tilde, is given by
\beq \lab{sdemod}
\d \tilde X = \Pe \, u(\tilde Y) \d t + \sqrt{2} \d W_1, \quad
\d \tilde Y = r(\tilde Y) \d t + \sqrt{2} \d W_2,
\eeq
instead of \eqn{sdeshear}. Here $r(y)$ is a function chosen so that the distribution of $\tilde Y$ better samples the regions where $u(y)$ is large (or small) which control $g(\xi)$ for $\xi$ away from $\xi_*$. Girsanov's formula relates averages under the original dynamics \eqn{sde} to averages under this modified dynamics according to
\[
\E \cdot = \tilde{\E} \cdot \e^{-\frac{1}{\sqrt{2}}\int_0^t r(\tilde Y(t')) \, \d W_2 - \frac{1}{4} \int_0^t r^2(\tilde Y(t')) \, \d t'}
\]
\citep{mils95,okse98}.
Thus $C(x,t)$ can be approximated by integrating numerically \eqn{sdemod} for an ensemble of trajectories and using a discretised version of the relation
\[
C(x,t) = \tilde{\E} \delta(x-\tilde X(t)) \e^{-\frac{1}{\sqrt{2}}\int_0^t r(\tilde Y(t')) \, \d W_2 - \frac{1}{4} \int_0^t r^2(\tilde Y(t')) \, \d t'}.
\]
This result is used for to estimate the tails of $C$ and hence the form of $g$ for large $|\xi|$ with a much better sampling than achieved with the original dynamics.
For the numerical results reported in \S\ref{sec:couette}--\ref{sec:poiseuille}, we have used $r(y)=\gamma (1-y)$ to efficiently sample the portion of $C(x,t)$ controlled by trajectories that remain localised near the wall at $y=1$ (leading to anomalously large $x$ for Couette flow and anomalously small $x$ for Poiseuille flow), and $r(y)=-\gamma y$ to sample trajectories localised near the maximum of the plane Poiseuille flow.  
The value of the parameter $\gamma$ was chosen by trial-and-error to obtain the best representation of a portion of the curve $g(\xi)$. A similar modified dynamics for both $Y(t)$ and $Z(t)$ was used in the case of the pipe Poiseuille flow in \S\ref{sec:pipe}.

\section{Small-$\Pe$ form of $f(\bq)$ for cellular flow} \label{sec:smallPe}

In the limit $\Pe \to 0$, the eigenvalue problem \eqn{eig1} can be solved perturbatively by introducing the expansions
\[
\phi = \phi_0 + \Pe \phi_1 + \Pe^2 \phi_1 + \cdots \inter{and}
f= f_0 + \Pe f_1 + \Pe^2 f_2 + \cdots
\]
of the eigenfunctions and eigenvalue into \eqn{eig1}. The leading-order, $O(1)$, equation is solved for $\phi_0=1$ and $f_0=|\bq|^2$ which reduces the $O(\Pe)$ equation to
\[
\nabla^2\phi_1 - 2 \bq \cdot \nabla \phi_1 + \bu \cdot \bq = f_1.
\]
On integrating over a period, the left-hand side vanishes, leading to $f_1=0$. The solution is then found in the form
\beq \lab{phi1smallpe}
\phi_1 = a \sin x \sin y + b \sin x \cos y + c \cos x \sin y + d \cos x \cos y,
\eeq
where the constants $a,\, b,\, c$ and $d$ are readily computed. Integrating the $O(\Pe^2)$ equation
\[
\nabla^2\phi_2 - 2 \bq \cdot \nabla \phi_2 - \bu \cdot \nabla \phi_1+ \bu \cdot \bq \, \phi_1 = f_2
\]
over a period leads to the eigenvalue correction
\[
f_2 = \frac{1}{(2\pi)^2} \int_0^{2\pi} \int_0^{2\pi} \left( -\bu \cdot \nabla \phi_1+ \bu \cdot \bq \, \phi_1\right) \, \d x \d y.
\]
Substituting \eqn{phi1smallpe} and taking the explicit form of the constants into account yields
\beq \lab{f2smallpe}
f_2 = \frac{1}{8} \frac{q_1^2 + q_2^2 + q_1^4 + 6 q_1^2 q_2^2 + q_2^4}{1+2(q_1^2+q_2^2) + (q_1^2-q_2^2)^2}.
\eeq

\bibliographystyle{agsm}

\end{document}